\documentclass[10pt]{elsarticle}
\usepackage{fullpage}

\usepackage[utf8]{inputenc}
\usepackage{amsmath,amsfonts}
\usepackage[mathscr]{euscript}
\usepackage{graphicx}
\usepackage[version=4]{mhchem}
\usepackage{siunitx}
\usepackage{longtable,tabularx}
\usepackage{booktabs}
\usepackage{bm}
\usepackage{subcaption}
\usepackage{tikz}
\usetikzlibrary{decorations.pathreplacing}
\usepackage{indentfirst}

\usepackage{hyperref,cleveref}
\makeatletter
\def\ps@pprintTitle{%
   \let\@oddhead\@empty
   \let\@evenhead\@empty
   \def\@oddfoot{\reset@font\hfil\thepage\hfil}
   \let\@evenfoot\@oddfoot
}
\makeatother

\newtheorem{remark}{Remark}

\begin{document}
\begin{abstract}
Cable subsystems characterized by long, slender, and flexible structural elements are featured in numerous
engineering systems. In each of them, interaction between an individual cable and the surrounding fluid is inevitable. 
Such a Fluid-Structure Interaction (FSI) has received little attention in the literature, possibly due to the inherent 
complexity associated with fluid and structural semi-discretizations of disparate spatial dimensions. This paper proposes 
an embedded boundary approach for filling this gap, where the dynamics of the cable are captured by a standard 
finite element representation $\mathcal C$ of its centerline, while its geometry is represented by a discrete surface 
$\Sigma_h$ that is embedded in the fluid mesh. The proposed approach is built on master-slave kinematics between 
$\mathcal C$ and $\Sigma_h$, a simple algorithm for computing the motion/deformation of $\Sigma_h$ based on the dynamic 
state of $\mathcal C$, and an energy-conserving method for transferring to $\mathcal C$ the loads computed on $\Sigma_h$. 
Its effectiveness is demonstrated for two highly nonlinear applications featuring large deformations and/or 
motions of a cable subsystem and turbulent flows: an aerial refueling model problem, and a challenging supersonic 
parachute inflation problem. The proposed approach is verified using numerical data, and validated using real flight data.
\end{abstract}


\begin{frontmatter}

\title{An Embedded Boundary Approach for Resolving the Contribution of Cable Subsystems to Fully Coupled Fluid-Structure Interaction}

\author[rvt1]{Daniel~Z.~Huang}
\ead{zhengyuh@stanford.edu}

\author[rvt3]{Philip~Avery}
\ead{pavery@stanford.edu}

\author[rvt1,rvt2,rvt3]{Charbel~Farhat}
\ead{cfarhat@stanford.edu}

\address[rvt1]{Institute for Computational and Mathematical Engineering,
               Stanford University, Stanford, CA, 94305}
\address[rvt2]{Mechanical Engineering, Stanford University, Stanford, CA, 94305}
\address[rvt3]{Aeronautics and Astronautics, Stanford University, Stanford, CA, 94305}

\end{frontmatter}

\section{Introduction}

Cable subsystems appear in a wide range of engineering systems and scientific applications, such as suspension
lines of parachutes and other atmospheric decelerators~\cite{fan2014simulation, tezduyar2010space, huang2018simulation,
sengupta2009supersonic, gao2016numerical, kim20093}, offshore drilling and production
risers~\cite{jaiman2009fully, holmes2006simulation, herfjord1999assessment}, and booms of airborne refueling 
systems~\cite{lofthouse2017cfd, ro2010modeling, zhu2007modeling, styuart2011numerical}. When immersed in a flow,
cable structures can be responsible for strong Fluid-Structure Interactions (FSIs) that may significantly affect
the performance of the system to which they are attached. For example, in the case of a dynamic, supersonic parachute
inflation, FSIs involving the suspension lines give rise to a significant drag reduction due to the disturbance of the front bow 
shock, while FSIs associated with marine drilling risers can lead to vortex-induced vibrations and fatigue damage of offshore systems.
Consequently, the ability to accurately and affordably model such phenomena has the potential to become a valuable tool
for the design and maintenance of the aforementioned and related engineering systems.

In the context of an individual cable, capturing a local, two-way, FSI interaction requires in general the accurate prediction of:
the flow-induced generalized forces and moments on the FE representation of the cable; and the effects of the structural dynamic response of the cable 
on the nearby fluid flow. Several efforts have been made to account for one or both of these two requirements. Strip theory, where computations 
are conducted on several cross sections, has been applied to the analysis of vortex-induced vibrations of deepwater 
risers~\cite{holmes2006simulation,herfjord1999assessment}. Empirical flow-induced load formulae and tabulated drag coefficients 
were adopted in~\cite{tezduyar2010space} for computing one-way FSIs associated with parachute suspension lines. 
In~\cite{kim20093, kim20062}, the effect of suspension lines on the fluid flow was modeled using source terms 
based on the inertial and elastic forces generated by the structure. By contrast, a brute force approach based on a body-fitted mesh 
that resolves the boundary layer of a dynamic cable was used in~\cite{jaiman2009fully} to successfully capture complex FSI phenomena 
driven by the cable. However: the computational cost of this nonadaptive approach can be overwhelming, particularly when the size of 
the cross section of the cable is small compared to its characteristic length; and the robustness of its underlying mesh motion
algorithm needed for maintaining at all times the Computational Fluid Dynamics (CFD) mesh body-fitted is usually limited to 
small-amplitude dynamics.

In addition to the computational complexity issue mentioned above for the case of the brute force approach adopted
in~\cite{jaiman2009fully}, the accurate prediction of a local, two-way coupled, FSI associated with a cable subsystem presents
many other challenges. First and foremost, the task of transferring information between fluid and structural meshes in order to
discretize the governing fluid-structure transmission conditions is complicated by the fact that in structural dynamics, the Finite
Element (FE) representation of a cable is typically carried out using line elements, and therefore is topologically One-Dimensional
(1D). To address this issue, a ``dressing'' approach based on ``phantom'' surface elements and massless rigid elements was developed by the third author almost
two decades ago~\cite{Vienna}. This noteworthy approach is equally applicable in both contexts of body-fitted~\cite{geuzaine2003aeroelastic} and non body-fitted CFD meshes~\cite{huang2018simulation}: 
however, it has computational disadvantages that are identified and discussed in this paper. For this reason, an alternative approach is proposed here for 
computing cable-driven FSI. This new approach is appropriate for the solution of both one-way and two-way FSI problems, is far more 
robust and computationally efficient than the brute force approach adopted in~\cite{jaiman2009fully}, and is more comprehensive
and user-friendly than the aforementioned dressing approach. 
It is based on: master-slave kinematics between the discrete line representing the 
centerline of a cable typically found in FE structural models, and a discretization of its true surface that is embedded 
(or immersed) in the computational fluid domain; an accurate algorithm for computing the displacement of the discrete surface of 
the cable based on the displacement and rotation of its centerline; and an energy-conserving method for distributing the 
flow-induced forces and moments on the nodes of the FE representation of the cable, based on the Degrees Of Freedom (DOFs) of this 
representation.

Another challenge for cable-driven FSI is the treatment of the potentially large motions and/or deformations of the cable, due its 
potentially large length-to-diameter ratio and/or high intrinsic flexibility. In the general case, this challenge is best addressed
by adopting the Eulerian computational framework for FSI which calls for an Embedded Boundary Method 
(EBM, e.g.~\cite{wang2011algorithms, farhat2012fiver}) -- also known as an Immersed Boundary Method (IBM, 
e.g.~\cite{peskin1977numerical, choi2007immersed}), or a Ghost Fluid Method (GFM, e.g.~\cite{tseng2003ghost, liu2006modified}) --
for CFD and FSI. Here, the Finite Volume method with Exact two-material Riemann problems (FIVER) -- which itself
is an embedded boundary framework for multi-material computations rather than a mere EBM for CFD and 
FSI~\cite{farhat2012fiver, lakshminarayan2014embedded, wang2015computational, main2017enhanced, huang2018family} --
is the chosen EBM. Specifically, both the dressing approach and the alternative master-slave kinematic approach 
proposed in this paper are incorporated into the Eulerian computational framework FIVER, and the Adaptive Mesh Refinement (AMR) 
method for EBMs described in~\cite{borker2019mesh} is applied to track the boundary layers around the cable subsystem and keep them
at all times well resolved.

The remainder of this paper is organized as follows. First, the dressing approach for cable-driven FSI problems is overviewed and 
discussed in~\Cref{sec:dressing approach}. Then, the alternative master-slave kinematic approach is introduced and analyzed 
in~\Cref{sec:alternative approach}. Next, the EBM for FSI in which both aforementioned approaches are incorporated is overviewed 
in~\Cref{sec:framework}, in order to keep this paper as self-contained as possible. In particular, small but effective novel 
contributions are made to this EBM and its AMR framework. The performance of the proposed approach for modeling cable-driven FSI is 
contrasted with that of the dressing approach in~\Cref{sec:apps}, first for an idealized airborne refueling system and then for a 
challenging, dynamic, supersonic parachute inflation problem. Finally, conclusions are offered in~\Cref{sec:conclusions}.

\section{Dressing approach for FSI models with disparate spatial dimensions}
\label{sec:dressing approach}

As already mentioned above, the structural dynamics of cable subsystems are generally modeled using topologically 1D beam elements, 
or special-purpose variants usually referred to as cable elements. This is due to their large length-to-diameter ratio. 
Semi-discretizations of cables based on beam elements have vastly superior computational efficiency than counterparts based on 
Three-Dimensional (3D) solid elements, provided that certain simplifying assumptions are appropriate -- for example, the assumption
that planar sections initially normal to the longitudinal axis of the cable remain planar and normal to this axis. On the other hand, 
a computational fluid domain is usually semi-discretized using 3D elements such as tetrahedra, hexahedra, and/or triangular prisms. 
Whether the FSI computations are performed using the Arbitrary Lagrangian-Eulerian (ALE) framework and a dynamic, body-fitted CFD 
mesh~\cite{hirt1974arbitrary, donea1982arbitrary, farhat2001discrete}, or the Eulerian 
framework, an embedded discrete surface and an EBM for CFD as well as
FSI~\cite{farhat2012fiver, lakshminarayan2014embedded, wang2015computational, main2017enhanced, huang2018family}, some difficulties 
if not challenges arise in the presence of this type of mixed semi-discretizations with disparate 
spatial dimensions. For instance, consider the case where the FSI computations are performed using the Eulerian framework, which is 
the more appropriate framework for highly nonlinear FSI problems characterized by large structural motions and/or deformations as 
well as topological changes. Within this computational framework, the detection of 1D structural elements embedded in the non
body-fitted CFD mesh and the definition as well as construction of associated geometric characteristics such as surface normals 
may be difficult, ambiguous, and/or computationally intensive. For this reason, capturing a cable-driven FSI using numerical 
techniques developed for other types of FSI problems characterized by convenient, Two-Dimensional (2D) fluid/structure interfaces 
is neither necessarily robust, nor necessarily computationally efficient, or even practical. 

\begin{figure}[!h]
  \centering
  \includegraphics[width=0.6\linewidth]{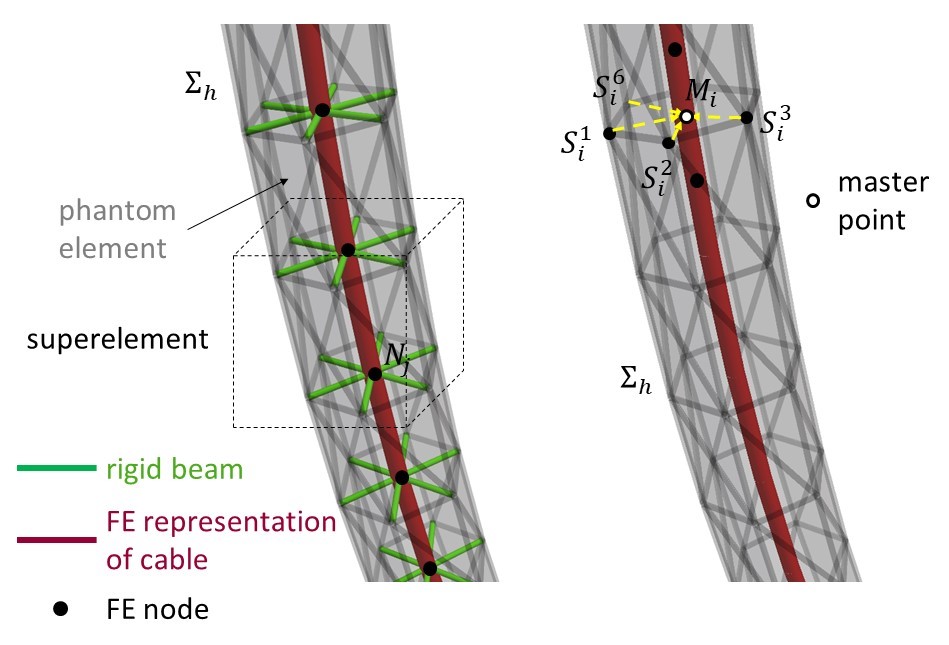}
  \caption{Schematics of the dressing approach based on dynamically equivalent superelements~(left), and counterpart schematics
	of the master-slave kinematic approach~(right): the discrete surface $\Sigma_h$ representing the true 
	geometry of the surface of the cable encloses the topologically 1D FE representation of the cable~(red); its nodes are 
	connected to the discretized cable by massless rigid elements (green) in the dressing approach (left), and by kinematics 
	constraints (yellow) in the master-slave kinematic approach (right); and $\{S_i^j\}_{j=1}^{n_i}$ denotes the set of $n_i$ 
	{\it coplanar} nodes in $\Sigma_h$ whose kinematics are slaved to those of the corresponding master {\it point} $M_i$ located 
	at the intersection of the discrete cross section defined by the set of nodes $\{S_i^j\}$ and the FE representation of the 
	cable.}
  \label{fig:concept}
\end{figure}

Developed initially for the so-called ``fish-bone'' aeroelastic computational models, the dressing approach first proposed
in~\cite{Vienna} for facilitating the enforcement of the fluid-structure transmission conditions at the interface
between fluid and structural semi-discretizations with disparate spatial dimensions can be described as a modeling approach based on 
the concept of a \emph{dynamically equivalent superelement} (see~\Cref{fig:concept}-left). In this approach, a superelement 
consists of:
\begin{itemize}
	\item A 1D flexible beam or cable element located at the centerline of the cable. 
	\item A cylindrical assembly of:
		\begin{itemize}
			\item A set of {\it massless}, {\it zero-stiffness}, surface elements of typical size $h$ -- referred to for 
				this reason as phantom elements -- and their nodes, located on the geometrically true surface of the 
				cable, $\Sigma$. 
			\item A network of {\it massless}, {\it rigid} beam elements connecting each FE node on the centerline of the cable 
				to the adjacent nodes of phantom elements. 
		\end{itemize}
\end{itemize}

Let $\Sigma_h$ and $\mathcal N$ denote the global set of phantom elements and global network of rigid beam elements, respectively.
From a structural dynamics viewpoint, each superelement defined above preserves the dynamics of the corresponding beam or cable element: in this 
sense, it is {\it dynamically equivalent} to this element. Indeed, this superelement has the same mass and mass distribution as the 
cable element it is associated with, because all elements in $\Sigma_h$ and those of $\mathcal N$ are massless. It has also the same 
stiffness and stiffness distribution because all elements in $\Sigma_h$ have zero stiffness and those in $\mathcal N$ are rigid. 
Unlike a beam or cable element however, this superelement is equipped with a representation of the true geometry of the 
surface of the cable where the FSI of interest occurs, which facilitates the loads evaluation process. These loads are transferred to 
the FE representation of the (flexible) cable via the rigid (beam) elements in $\mathcal N$.

Specifically, $\Sigma_h$ can be used in both contexts of body-fitted and non body-fitted CFD meshes to directly couple the 
computational fluid and structural domains at the fluid/structure interface using a matching (or pairing)
procedure such as that described 
in~\cite{farhat1998load}, which is commonly used for enforcing the semi-discrete transmission conditions across a non-conforming 
fluid/structure interface. The flow-induced loads can be conveniently evaluated on $\Sigma_h$ and transferred to the FE structural 
representation of the cable, and the kinematics of the cable (including time-derivatives) can be equally 
conveniently transferred to the computational fluid domain to update its wall boundary conditions. The massless rigid beam elements 
couple the DOFs of the discrete surface $\Sigma_h$ with those of the FE semi-discretization of the cable using algebraic constraint 
equations. These can be enforced using, for example, the Lagrange multiplier method, which transforms the overall governing
semi-discrete equations of equilibrium of the structural subsystem into a global, Differential-Algebraic Equation (DAE). 

The dressing approach outlined above can be used to construct computational FSI models for a variety of applications.
These include, for example, the aeroelastic analysis of highly flexible aircraft, rotorcraft, and turbine blades.
It is equally applicable to cable-driven FSI problems. Although tedious, the generation of the set of surface elements $\Sigma_h$ and 
the construction of the network of rigid elements $\mathcal N$ can be automated~\cite{CMS}. Nevertheless, this approach has a few 
noteworthy shortcomings, particularly for cable applications:
\begin{itemize}
	\item Its practical implementation requires a certain discipline. For example, $\Sigma_h$ and $\mathcal N$ are best constructed
		so that each DOF attached to each node of a phantom element in $\Sigma_h$ is connected to a corresponding DOF of
		a rigid element in $\mathcal N$, or otherwise to any other DOF of the FE structural model, in order to avoid the 
		generation of artificial mechanisms in the overall FE structural model -- that is, the introduction in this model of 
		DOFs with zero associated stiffness.
	\item It increases the number of structural DOFs by typically an order of magnitude. Given the computational cost associated 
		with modeling a turbulent flow problem, this issue may be relatively insignificant. However, it is by no means 
		desirable. 
	\item It leads to a mass matrix of the overall FE structural model that is singular. Specifically, this matrix has a zero 
		entry at each DOF attached to a node of an element in $\Sigma_h$, due to the massless nature of such an element as 
		well as the massless nature of any rigid beam element in $\mathcal N$ to which it is connected. Consequently, it 
		unnecessarily complicates the implementation of various computational modules of an FSI solver:
		\begin{itemize}
			\item For explicit time-integration schemes requiring a positive definite mass matrix -- which are commonly 
				used for highly nonlinear FSI problems involving contact and/or crack propagation 
				(e.g., see~\cite{wang2015computational}) -- it necessitates the implementation of a cumbersome static 
				condensation in the main time-integration loop of the structural analyzer.
			\item For implicit time-integration schemes, the constraint equations underlying the rigid elements in 
				$\mathcal N$ introduce a destabilizing effect in the overall FE structural dynamics 
				subsystem~\cite{farhat1995implicit}, due to its DAE aspect. The mitigation of this effect requires 
				either the introduction in the time-integrator of artificial damping, which lowers its order of 
				time-accuracy, or a special purpose formulation of the structural dynamics equations of equilibrium, 
				which complicates its numerical implementation \cite{farhat1995implicit}. 
		\end{itemize} 
\end{itemize}

By contrast, the alternative approach presented next for facilitating the enforcement of fluid-structure transmission 
conditions at an interface between fluid and structural semi-discretizations with disparate spatial dimensions does not suffer
from any such pitfalls.

\section{Alternative master-slave kinematic approach}
\label{sec:alternative approach}

The issues associated with the dressing technique outlined above can be resolved through an alternative approach for achieving
the same objective, where (see Figure~\ref{fig:concept}-right):
\begin{itemize}
	\item A discrete representation $\Sigma_h$ of the surface of each cable is constructed as: a collection of sets of coplanar 
		nodes $\{S_i^j\}_{j=1}^{n_i}$, where each set defines a discrete representation of an $i$-th cross section of the 
		cable; and a set of surface elements that connect these nodes.
	\item $\Sigma_h$ is embedded in the given CFD mesh associated with the overall computational fluid domain.
	\item Three displacement DOFs are attached to each node $S_i^j \in \Sigma_h$.
	\item These DOFs are slaved to the translational as well as rotational motion of the cable at the point $M_i$
		located at the intersection of the discrete cross section defined by $\{S_i^j\}_{j=1}^{n_i}$ and the FE 
		representation $\mathcal C$ of the cable ($M_i$ is also the closest point in $\mathcal C$ to any node 
		$S_i^j \in \Sigma_h$).
	\item The translational as well as rotational motion of the cable at each point $M_i$ is computed by interpolation of the 
		displacement and rotational DOFs attached to the nodes of the FE beam or cable element $e_i$ containing the point 
		$M_i$.
\end{itemize}
\begin{remark}The notation $\Sigma_h$ adopted in the dressing approach for describing the discrete representation of the surface of a cable
is reused here because the same discretization of the cable's surface can be reused in the alternative approach described
herein. However, the dynamics of $\Sigma_h$ differ in both approaches.
\end{remark}

As in the dressing approach, the CFD mesh associated with the overall computational fluid domain can be in this case body-fitted or 
non-body fitted. However, given the intrinsic advantages of the Eulerian framework for FSI problems characterized by \emph{large} 
structural motions and/or deformations, attention is focused here on the case of a non body-fitted mesh -- and therefore on the case
of a preferred EBM for CFD and FSI. Furthermore, for simplicity but without any loss of generality, the structural subsystem is 
assumed to consist of a single cable: this simplifies the description of the proposed method, but in no way limits its scope
of applications to a single cable or an assembly of cables.

In the alternative approach described here, the chosen structural analyzer computes at each time-step the state of the FE
representation $\mathcal C$ of the cable subsystem. However, it is not given direct access to the preferred flow solver. Similarly,
the flow solver is not given direct access to the computed state of the discretized cable subsystem. Instead, the kinematics of 
$\Sigma_h$ -- that is, the position and velocity of this embedded discrete surface -- are slaved to the computed state of the FE
representation $\mathcal C$ of the cable, and the flow-induced loads are computed on $\Sigma_h$ before they are transferred to 
$\mathcal C$. Specifically, these motion and loads transfers are performed as follows:

\begin{itemize}
	\item Preprocessing step:
		\begin{itemize}
			\item Pair each slave node $S_i^j \in \Sigma_h,~ j = 1,~\cdots,~n_i$, with the closest element 
				$e_i \in \mathcal C$ containing the master point $M_i$. (Note that the superscript $j$ highlights the 
				surjective aspect of the function $S_i^j \longrightarrow M_i$: specifically, $n_i \ge 3$ nodes $S_i^j$
				are typically paired with one point $M_i$). The initial distance vector $\bm{d}_i^{j}$ between these 
				two locations -- that is, the distance vector at $t = t^0$ -- is defined by
				\begin{equation*} 
					\bm{x}^0_{S^j_i} = \bm{x}^0_{M_i} + \bm{d}_i^{j} 
					\Leftrightarrow \bm{d}_i^j = \bm{x}^0_{S^j_i} - \bm{x}^0_{M_i}
				\end{equation*}
				where here and throughout the remainder of this paper, the bold font designates a vector quantity and
				$\bm{x}_P$ denotes the position vector of a point $P$. Consistently with the modeling assumptions of a
				FE beam element, the distance vector $\bm{d}_i^j$ is assumed to be time-independent.
		\end{itemize}
	\item At each time-instance $t^n$: 
		\begin{itemize}
		
			\item Compute the displacement $\bm{u}_{S_i^j}^n$ and velocity $\dot{\bm{u}}_{S_i^j}^n$ of the slave node 
				$S_i^j$ as follows
				\begin{equation} 
					\bm{u}_{S_i^j}^n = \bm{u}_{M_i}^n + \bm{R}(\bm{\theta}_{M_i}^n)\, \bm{d}_i^{j} - \bm{d}_i^{j}
					~~~~~~\textrm{and}~~~~~~~~ 
					\dot{\bm{u}}_{S_i^j}^n = \dot{\bm{u}}_{M_i}^n 
					+ \bm{\omega}_{M_i}^n \times \bm{R}(\bm{\theta}_{M_i}^n)\,\bm{d}_i^{j}
					\label{eq:disp} 
				\end{equation}
				where: the dot designates the time-derivative; $\bm{u}_{M_i}^n$ and $\bm{\theta}_{M_i}^n$ denote the interpolated displacement and rotation
				vectors at the point $M_i$, respectively; $\dot{\bm{u}}_{M_i}^n$ and ${\bm{\omega}}_{M_i}^n$ are the 
				interpolated velocity and angular velocity vectors at the point $M_i$, respectively; and $\bm{R}$
				is the rotation matrix at $M_i$ and depends on $\bm{\theta}_{M_i}^n$.  
			\item Compute the force vector $\bm{f}^n_{S^j_i}$ at each slave node $S_i^j$ as follows 
			\begin{equation}
 				 \bm{f}^n_{S^j_i} = \int_{\Sigma_h^n} \Big( -p^n \bm{n}^n + \bm{\tau}^n \bm{n}^n\Big)\phi_{S^j_i}\,d\Sigma_h
				 \label{eq:fs}
			\end{equation}
				where: $p^n$ and $\bm{\tau}^n$ denote the pressure and viscous stress tensor of the flow at time $t^n$;
				$\bm{n}^n$ denotes the outward normal to $\Sigma_h^n$ at time $t^n$;  and $\phi_{S^j_i}$ denotes a {\it local} shape 
				function associated with the node $S^j_i \in \Sigma_h$ (for example, the characteristic function
				at node $S_i^j$ in the case of a finite volume semi-discretization of the flow equations). 
				In practice, \Cref{eq:fs} is evaluated using a quadrature rule in each surface 
				element~(see \Cref{sec:load_comp}).
			\item Compute the  force vector $\bm{f}^n_{M_i}$ and moment vector $\bm{m}^n_{M_i}$ at the point $M_i$ 
				as follows
				\begin{equation} 
					\bm{f}^n_{M_i} = \sum_{j = 1}^{n_i}\bm{f}^n_{S^j_i}  ~~~~~~~\textrm{and}~~~~~~~~
					\bm{m}^n_{M_i} = \sum_{j = 1}^{n_i}\bm{R}(\bm{\theta}^n_{M_i})\,\bm{d}^j_i 
					\times \bm{f}^n_{S^j_i}
					\label{eq:load_M} 
				\end{equation} 
			\item Apply the load transfer method presented in~\cite{farhat1998load} to deduce the following
				expressions of the generalized force and moment vectors acting on a FE node $N_j$ of $\mathcal C$, 
				$\bm{f}_{N_j}^n$ and $\bm{m}_{N_j}^n$, respectively
				\begin{equation} 
					\bm{f}^n_{N_j} = \sum_{i: N_j \in E_i}\bm{f}_{M_i}^n \phi_{N_j}^d (M_i) 
					~~~~~~~\textrm{and}~~~~~~~~
					\bm{m}^n_{N_j} = \sum_{i: N_j \in E_i}\bm{m}_{M_i}^n \phi_{N_j}^r (M_i)
					\label{eq:load_N} 
				\end{equation} 
				where $\phi_{N_j}^d$ and $\phi_{N_j}^r$ denote the standard shape functions associated with 
				the displacement and rotational DOFs at node $N_j \in \mathcal C$, respectively.
		\end{itemize}
\end{itemize}

The master-slave kinematic approach described above for enforcing the fluid-structure transmission conditions at an interface between fluid 
and structural semi-discretizations with disparate spatial dimensions effectively decouples the DOFs of the slave nodes from the 
structural analyzer, thereby avoiding zero mass and other destabilizing singularities that may arise in the dressing approach and 
reducing computational overhead. Yet, despite this decoupling, this alternative approach is globally conservative as shown below. 

Consider at time $t^n$ a virtual fluid displacement field $\delta\bm{u}^n_F$ of the fluid/structure boundary $\Sigma_h$. The 
corresponding virtual work performed by the fluid tractions acting on $\Sigma_h$ can be written as
\begin{equation}
	-\delta {\mathcal W}^n_F = \int_{\Sigma_h^n} \Big( -p^n \bm{n}^n + \bm{\tau}^n \bm{n}^n\Big)\cdot\delta\bm{u}^n_F\, d\Sigma_h = \sum_{i=1}^{n_M} \sum_{j = 1}^{n_i}\bm{f}^n_{S^j_i} \cdot \delta \bm{u}^n_{S^j_i}
\label{eq:work_f}
\end{equation}
where: $n_M$ denotes the number of discrete cross sections of the cable represented in $\Sigma_h$ -- or equivalently, the number of master points 
$M_i$; $\bm{f}_{S^j_i}$ is given in \Cref{eq:fs}; and the no-slip displacement transmission condition (viscous flow) is assumed to
hold ($\delta \bm{u}^n_F = \delta \bm{u}^n_S$). 

Let $n_N$ denote the number of FE nodes in $\mathcal C$. From~\Cref{eq:work_f}, \Cref{eq:disp}, \Cref{eq:load_M}, and \Cref{eq:load_N},
it follows that
\begin{equation*}
  \begin{split}          
	  -\delta {\mathcal W}^n_F &= \sum_{i=1}^{n_M} \sum_{j=1}^{n_i} \bm{f}^n_{S^j_i} \cdot \Big(\delta \bm{u}^n_{M_i} 
                 + \delta\bm{\theta}^n_{M_i}\times \bm{R}(\bm{\theta}^n_{M_i})\, \bm{d}^j_i\Big)\\
              &= \sum_{i=1}^{n_M} \sum_{j=1}^{n_i} \bm{f}^n_{S^j_i} \cdot \delta \bm{u}^n_{M_i}
                 + \Big(\bm{R}(\bm{\theta}^n_{M_i})\, \bm{d}^j_i \times \bm{f}^n_{S^j_i}\Big) \cdot \delta\bm{\theta}^n_{M_i}\\
              &= \sum_{i=1}^{n_M} \bm{f}^n_{M_i} \cdot \delta \bm{u}^n_{M_i} + \bm{m}^n_{M_i} \cdot \delta\bm{\theta}^n_{M_i}\\
              &= \sum_{i=1}^{n_M} \bm{f}^n_{M_i} \cdot \Big(\sum_{N_j\in E_i} \phi_{N_j}(M_i) \delta \bm{u}^n_{N_j} \Big)+ 
	      \bm{m}^n_{M_i} \cdot \Big(\sum_{N_j\in E_i} \phi_{N_j}(M_i)  \delta\bm{\theta}^n_{N_j}\Big)\\
              &= \sum_{j=1}^{n_N} \bm{f}^n_{N_j} \cdot \delta \bm{u}^n_{N_j} + \bm{m}^n_{N_j} \cdot \delta\bm{\theta}^n_{N_j}\\
	      &= \delta {\mathcal W}^n_S
  \end{split}
\end{equation*}
which shows that the master-slave kinematic approach proposed here for performing motion and load transfers in cable-driven FSI is 
globally conservative.

\begin{remark}The master-slave kinematic approach described above is independent of the shape of the cross section of the cable, albeit this shape affects the construction of the embedded discrete surface 
$\Sigma_h$.
\end{remark}

\section{Eulerian framework for fluid-structure interaction}
\label{sec:framework}

Again, both dressing and master-slave kinematic approaches for enabling the simulation of cable-driven FSI problems 
characterized by disparate spatial dimensions of the computational fluid and structural domains can work with body-fitted as well as 
non body-fitted CFD meshes. However, because long, slender cables tend to be highly flexible and therefore tend to undergo large 
motions and/or deformations, the focus of this paper is set on non body-fitted CFD meshes and therefore on the Eulerian framework for 
FSI equipped with an EBM. Since all FSI computations reported in this paper are performed using the second-order EBM
FIVER~\cite{wang2011algorithms, farhat2012fiver, lakshminarayan2014embedded, wang2015computational, main2017enhanced, huang2018family},
this Finite Volume (FV)-based EBM is overviewed below in order to keep this paper as self-contained as possible. Specifically, 
attention is focused on the case where the embedded discrete surface $\Sigma_h$ exclusively emanates from the representation 
of the surface of a cable in the master-slave kinematic approach described in \Cref{sec:alternative approach}. For the case of an 
arbitrary embedded discrete surface $\Sigma_h$ -- and therefore the case of more general fluid-structure transmission conditions 
than~\Cref{eq:disp} and~\Cref{eq:load_N} -- and/or for further details on the FIVER framework, the reader is referred to the 
aforementioned references.

\subsection{The second-order embedded boundary method FIVER for fluid-structure interaction}

Consider a compressible, turbulent, viscous flow governed by the Navier-Stokes equations with turbulence modeling. If these equations 
are written in conservation form, their semi-discretization by a second-order, vertex-based, FV method leads to the following system 
of ordinary differential equations
\begin{equation}
	\bm{V} \dot{\bm{W}} + \bm{F}(\bm{W}) - \bm{G}(\bm{W}, \bm{W}_t) = \bm{0}
\label{eq:SDF}
\end{equation}
where $\bm{W}$ denotes the vector of semi-discrete, conservative, fluid state variables, $\bm{V}$ is the diagonal matrix storing the
dual cells (or control volumes), $\bm{F}$ denotes the vector of numerical convective fluxes, $\bm{G}$ denotes the vector of numerical 
diffusive fluxes and source terms due to turbulence modeling, and $\bm{W}_t$ denotes the semi-discrete state vector of the turbulence 
model conservation variables.

\subsubsection{Numerical convective fluxes} 

\begin{figure} [!h]
	\centering 
	\includegraphics[width=0.7\linewidth]{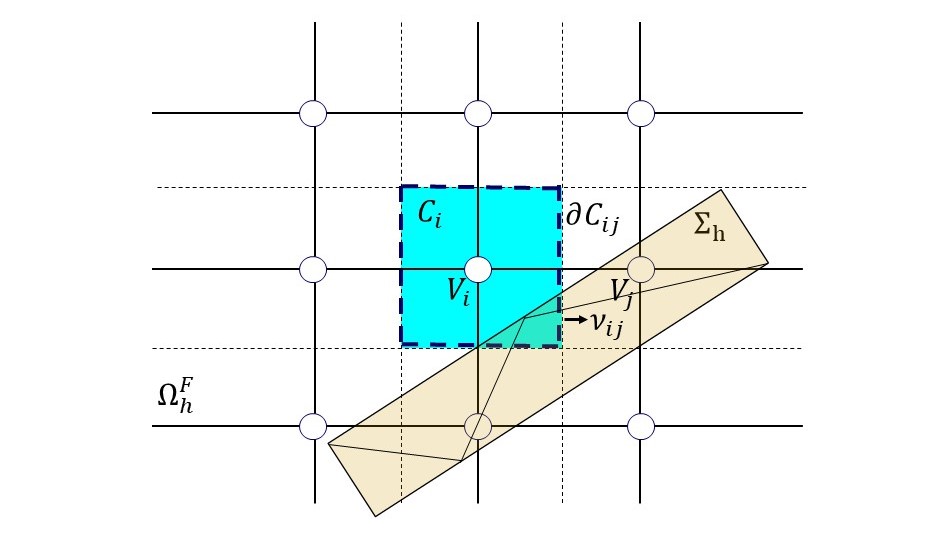}
	\caption{Discretization $\Omega^F_h$ of an embedding fluid domain, dual cell (control volume) $C_i$, boundary facet 
	$\partial C_{ij}$, unit outward normal $\bm{\nu}_{ij}$, and embedded discrete surface $\Sigma_h$ (two-dimensional case).}
	\label{fig:fluid_domain}
\end{figure}

Typically, the numerical convective fluxes are computed in a vertex-based FV method on an edge-by-edge basis. Let $\Omega^F$ denote 
the embedding computational fluid domain, $\Omega^F_h$ denote its discretization by a structured or unstructured, non body-fitted,
CFD mesh with dual cells $C_i$, and $\Omega_h^{\Sigma_h}$ denote the restriction of $\Omega_h^F$ to the volume enclosed by the 
embedded discrete surface $\Sigma_h$ representing here the geometrically true surface of the cable (see~\Cref{fig:fluid_domain}). Let 
also $V_i$ denote a generic node of the embedding CFD mesh, $K(V_i)$ denote the set of nodes $V_j$ of this mesh that are connected by 
an edge $V_iV_j$ to $V_i$, and $\bm{\nu}_{ij}$ denote the unit outward normal to the control volume boundary facet $\partial C_{ij}$ 
connecting the centroids of the elements of the embedding CFD mesh sharing the nodes $V_i$ and $V_j$ (see~\Cref{fig:fluid_domain}).

At any node $V_i$ away from $\Sigma_h$ -- that is, any node where $\forall V_j \in K(V_i), \, V_iV_j \bigcap \Sigma_h = \{\emptyset\} 
\Leftrightarrow  V_iV_j \in \Omega_h^F \backslash \Omega_h^{\Sigma_h}$ -- a second-order, vertex-based, FV method computes the vector 
of numerical fluxes $\bm{F}_i$ as follows
\begin{equation*}
	\bm{F}_i = \displaystyle\sum_{\substack{V_j\in K(V_i)\\  V_iV_j \bigcap \Sigma_h = \{\emptyset\}}} \bm{\Phi}_{ij}(\bm{W}_i, 
	\bm{W}_j, \bm{\nu}_{ij}, \hbox{EOS})
\end{equation*}
where $\bm{\Phi}_{ij}$ denotes a numerical flux vector function associated, for example, with a second-order extension of a 
first-order upwind scheme -- such as Roe's approximate Riemann solver~\cite{roe1981approximate} -- based on the Monotonic Upwind 
Scheme Conservation Law (MUSCL)~\cite{van1979towards}, $\bm{W}_i$ and $\bm{W}_j$ denote the semi-discrete fluid state vectors at the
nodes $V_i$ and $V_j$, respectively, and EOS refers to the Equation of State characterizing the considered compressible fluid.

\begin{figure}[!h]
\centering 
	\includegraphics[width=0.6\linewidth]{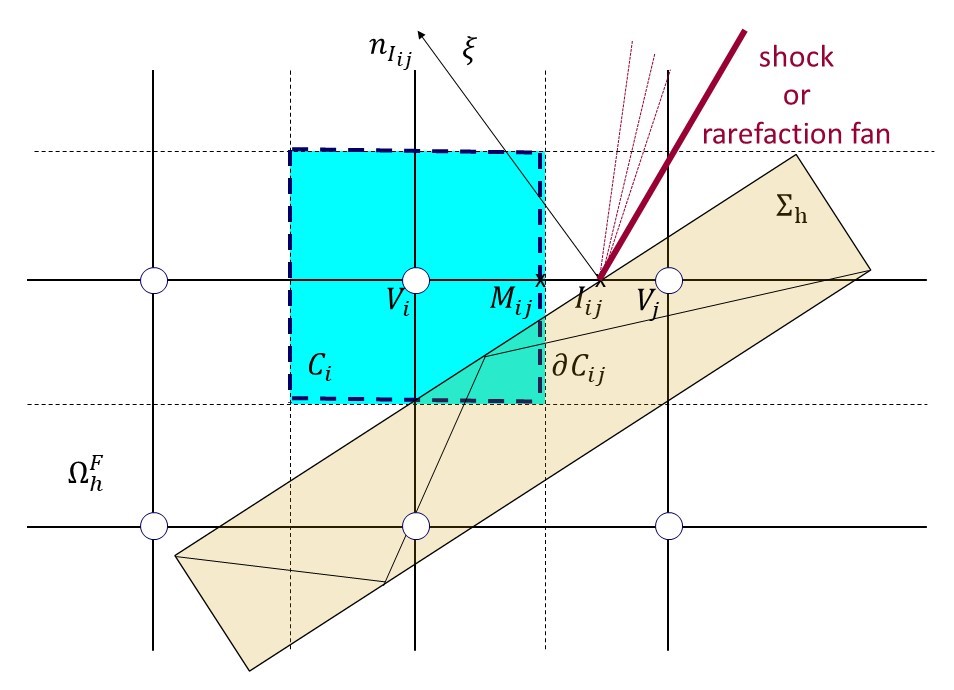} 
	\caption{Construction and solution of a local, 1D, exact, fluid-structure half Riemann problem at the fluid/structure material interface (two-dimensional case).}
  \label{fig:HRF}
\end{figure}

In the vicinity of the embedded discrete surface $\Sigma_h$ defined here by $V_iV_j \bigcap \Sigma_h \ne \{\emptyset\}$, a ``mixed 
(dual) cell'' problem arises. For example, if $V_i \in \Omega_h^F \backslash \Omega_h^{\Sigma_h}$ but $V_iV_j \bigcap \Sigma_h \ne 
\{\emptyset\}$ -- or in other words, $V_i \in \Omega_h^F \backslash \Omega_h^{\Sigma_h}$ but $V_j \in \Omega_h^{\Sigma_h}$ -- the 
computational dual cell $C_i$ attached to $V_i$ is in this case a mixed cell as it is partly occupied by the fluid and partly occupied
by the structure (see~\Cref{fig:fluid_domain} and~\Cref{fig:HRF}): such a situation complicates the semi-discretization process in 
$C_i$, and in particular the evaluation of the numerical convective flux vector function $\bm{\Phi}_{ij}$ across the boundary facet 
$\partial C_{ij}$. The second-order EBM FIVER~\cite{main2017enhanced, huang2018family} addresses this issue by reconstructing the 
semi-discrete fluid state vector $\bm{W}_i$ at $\Sigma_h$ and solving there a 1D, exact, fluid-structure, half Riemann problem 
designed to correctly semi-discretize the convective flux in a mixed cell. At each time-step $t^n$, this can be written as described 
below, with all superscripts referring to the time-instance $t^n$ omitted except in $t^n$ and $t^{n+1}$ in order to keep the notation 
as simple as possible:

\begin{enumerate}
	\item Compute a linear reconstruction of the vector of semi-discrete, primitive, fluid state variables at the intersection 
		point $I_{ij}$ of $V_iV_j$ and $\Sigma_h$ 
\begin{equation}
	\bm{w}_{I_{ij}} = \bm{w}_i + \bm{\nabla w}_i\cdot(\bm{x}_{I_{ij}} - \bm{x}_i)
	\label{eq:recons}
\end{equation}
		where $\bm{w} = (\rho, \bm{v}, p)$, and $\rho$, $\bm{v}$, and $p$ denote the fluid density, velocity vector, and pressure, respectively.
\item Construct and solve in the time-interval $[t^n, t^{n+1}]$, at the moving point $I_{ij}$ and along the axis $\vec \xi$ whose origin is the moving point $I_{ij}$
	and direction is the normal $\bm{n}_{I_{ij}}$ to the moving material interface $\Sigma_h$ at $I_{ij}$, the following local, 1D, exact,
		fluid-structure, half Riemann problem 
		\begin{equation}
			 \begin{array} {r c l l} 
				 \displaystyle{\frac{\partial \bm{w}}{\partial t} + \frac{\partial}{\partial \xi} \left[\mathcal{F}\left(\bm{w}\right) - {\dot u}_{I{ij}} \bm{w}\right]} & = & 0, & 
				 \qquad \xi \in [0, \infty) \\ 
				 \bm{w}(\xi, 0) & = & \bm{w}_{I_{ij}}, & \qquad \xi \in [0, \infty)\\ 
				 \left(v_{I_{ij}} :=  \bm{v}_{I{ij}}\cdot\bm{n}_{I_{ij}}\right) (t-t^n) & = & 
				 \left(\dot{u}_{I{ij}} := \dot{\bm{u}}_{I{ij}}\cdot\bm{n}_{I_{ij}}\right) (t-t^n) & \qquad \forall t \in [t^n, t^{n+1}] 
			 \end{array} 
			 \label{eq:HRP} 
		\end{equation}
		where: $\bm{w}(\xi, t) = \left(\rho(\xi, t), (v(\xi, t) := \bm{v}(\xi, t)\cdot\bm{n}_{I_{ij}}), p(\xi, t)\right)$; 
		$\xi$ denotes the abscissa along the $\vec \xi$ axis defined 
		above; $\bm{w}_{I_{ij}} = (\rho_{I_{ij}}, v_{I_{ij}}, p_{I_{ij}})$ is the reconstructed vector of discrete, primitive, 
		fluid-state variables at the point $I_{ij}$ (see~\Cref{eq:recons}); $\mathcal F$ denotes the continuous convective flux vector; 
		and $\dot{\bm{u}}$ denotes the velocity vector of $\Sigma_h$ and is given by the second of 
		Equations~(\ref{eq:disp}). Assuming that $\dot{\bm{u}}_{I_{ij}}$ is constant within the time-interval $[t^n, t^{n+1}]$, the solution of 
		the above half Riemann problem can be derived analytically when the EOS of the fluid allows it, or otherwise efficiently computed 
		numerically using sparse grid tabulations~\cite{farhat2012fiver}. In either case, this solution consists of two constant states separated 
		in the $(\xi, t)$ plane by either a shock wave or a rarefaction fan~(see~\Cref{fig:HRF}). In practice, however, only the solution at the
		point $I_{ij}$, $\bm{w}^{\mathcal R} = \left(\rho_{I_{ij}}^{\mathcal R}, (v_{I_{ij}}^{\mathcal R} = \dot{u}_{I{ij}}), p_{I_{ij}}^{\mathcal R}\right)$, 
		needs be computed and stored. It is worthwhile noting that for an inviscid flow, solving the above half Riemann problem enforces the slip 
		transmission condition at the point $I_{ij}$ through the third  of Equations~(\ref{eq:HRP}), and computes the values of the flow density 
		and pressure at this point. For a viscous flow however, solving the half Riemann problem~(\ref{eq:HRP}) simply computes the fluid state 
		vector at the intersection point $I_{ij}$ of the edge $V_iV_j$ and $\Sigma_h$ in order to enable in Step 5 below the computation of the 
		numerical flux vector function $\bm{\Phi}_{ij}$ across the boundary facet $\partial C_{ij}$ of the mixed cell $C_i$ -- and hence, the 
		semi-discretization of the convective fluxes in a mixed cell. In both cases, it enables the numerical approximation to capture any shock or 
		rarefaction wave near the fluid/structure interface $\Sigma_h$.
	\item Update the discrete fluid velocity vector at the point $I_{ij}$ by replacing its reconstructed normal component by $v_{I_{ij}}^{\mathcal R}$ 
		$$\bm{v}_{I_{ij}}^{\star} =  v_{I_{ij}}^{\mathcal R} \bm{n}_{I_{ij}} + \left(\bm{v}_{I_{ij}} - (\bm{v}_{I_{ij}} 
		\cdot \bm{n}_{I_{ij}})\bm{n}_{I_{ij}}\right)$$ 
		and update the vector of semi-discrete, conservative, fluid state variables at the point $I_{ij}$ accordingly
		$$\bm{W}_{I_{ij}}^{\star} = \bm{W}(\rho_{I_{ij}}^{\mathcal R}, \bm{v}_{I_{ij}}^{\star}, p_{I_{ij}}^{\mathcal R})$$ 
	\item Compute the vector of semi-discrete, conservative, fluid state variables at the intersection of the edge $V_iV_j$ and the 
		boundary facet $\partial C_{ij}$ -- which is also the midpoint $M_{ij}$ along the edge $V_iV_j$ (see~\Cref{fig:HRF}) -- as follows
		\begin{equation} 
			\bm{W}_{ij} = \alpha \bm{W}_i + (1 - \alpha)\bm{W}_{I_{ij}}^{\star} 
			\label{eq:IOE}
		\end{equation}
		where $\alpha = \displaystyle{\frac{\|\bm{x}_{I_{ij}} - \bm{x}_{M_{ij}}\|_2}{\|\bm{x}_{I_{ij}} - \bm{x}_i\|_2}}$.
		Note that (\ref{eq:IOE}) implies an interpolation if $I_{ij}$ happens to be between $M_{ij}$ and $V_j$, or an extrapolation otherwise.
		Also, see \cite{huang2018family} for a stabilization of (\ref{eq:IOE}) when $I_{ij}$ is very close to $V_i$.
	\item Finally, compute the numerical flux vector function $\bm{\Phi}_{ij}$ across the boundary facet $\partial C_{ij}$ of the mixed cell $C_i$ 
		as follows
      \begin{equation}
	      \bm{\Phi}_{ij} = \bm{\Phi}_{ij}\left(\bm{W}_i, \bm{W}_{ij}, \bm{\nu}_{ij}, \hbox{EOS}\right)
	      \label{eq:finalF}
      \end{equation}
\end{enumerate}

In summary, the second-order FIVER method computes the vector of numerical convective fluxes $\bm{F}$ (see~\Cref{eq:SDF}) on a
non body-fitted mesh, away from an embedded discrete surface $\Sigma_h$, exactly as a standard, second-order, vertex-based, FV
method computes $\bm{F}$ on a body-fitted CFD mesh. In the vicinity of $\Sigma_h$ however, the second-order FIVER method computes
the contributions to $\bm{F}$ of each mixed cell as described in Equations~(\ref{eq:recons})--(\ref{eq:finalF}).

\begin{remark}From the above description of the semi-discretization of the convective fluxes, it follows that for inviscid FSI problems,
the second-order EBM FIVER -- and for this matter, any of the FIVER methods published so far --  does not introduce, exploit, or rely 
on the concept of a ghost flow, ghost boundary, or a ghost cell~(for example, see~\cite{laney1998computational}). This is in sharp 
contrast with most of the alternative EBMs, IBMs, and GFMs, including those cited in the introduction of this paper.
\end{remark}

\subsubsection{Numerical diffusive fluxes}

Like many FV-based methods, the second-order EBM FIVER -- and for this matter, any FIVER method published so far -- approximates the 
diffusive fluxes using a FE-like approach -- that is, on an element-by-element basis. For this reason, it computes the diffusive 
numerical fluxes on the primal cells of the embedding CFD mesh -- that is, on the ``elements'' of this mesh rather than its control 
volumes. 

In each primal cell of an embedding CFD mesh located away from the material interface $\Sigma_h$, the second-order EBM FIVER 
computes the contributions to the vector of numerical diffusive fluxes $\bm{G}$ exactly as a standard, second-order, vertex-based,
FV method computes these contributions on a body-fitted CFD mesh.

\begin{figure}
\centering
	\includegraphics[width=0.7\linewidth]{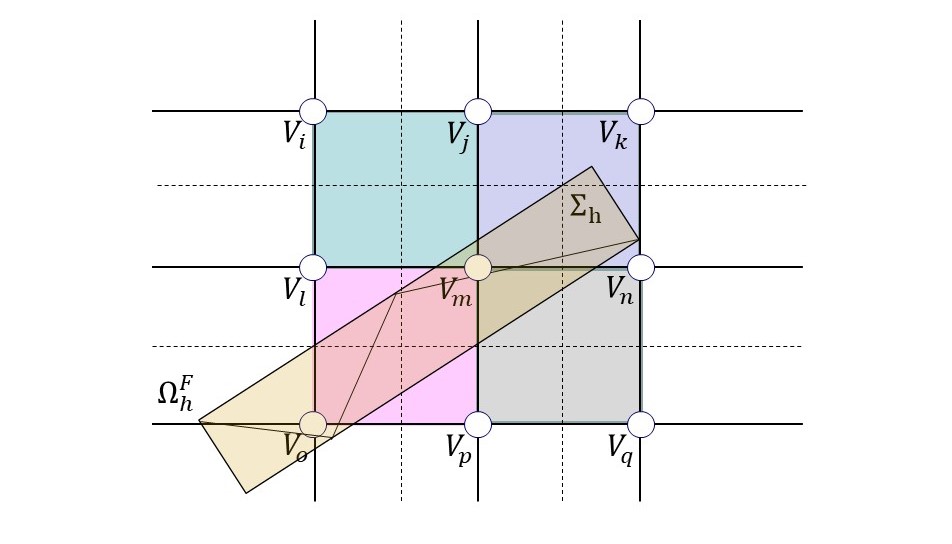} 
	\caption{Global approach for populating variables of the fluid state vector at the ghost fluid node $V_m$ involving 
	information extracted from the real nodes $V_i$, $V_j$, $V_k$, $V_l$, $V_n$, $V_p$, and $V_q$, vs. local approach featuring 4 
	different fluid state vectors at $V_m$, each computed using information extracted from 1 of the 4 elements connected to this
	node and to be used in the computation of the contributions of this element to the vector of numerical diffusive fluxes (two-dimensional case).}
\label{fig:GFM}
\end{figure}

In a mixed (primal) cell however, the second-order EBM FIVER distinguishes between ghost/real, (inactive/active, or 
occluded/unoccluded) fluid nodes. Essentially, it identifies first the ghost (a.k.a inactive or occluded) fluid nodes and populates
the velocity vector $\bm{v}$ and temperature $T$ at these nodes of the embedding CFD mesh using the standard mirroring 
technique~\cite{laney1998computational}, or a combination of constant and linear extrapolations from the neighboring real
(a.k.a. active or unoccluded) nodes~\cite{lakshminarayan2014embedded}. For example, consider the primal cell 
$(V_m, V_n, V_k, V_j)$ of the scenario illustrated in~\Cref{fig:GFM}. In this case, the second-order EBM FIVER identifies $V_m$ as an 
occluded fluid node in this primal cell. Then, it populates $\bm{v}$ and $T$ at this node using a combination of constant and 
linear extrapolations of their counterpart values at the unoccluded neighboring nodes $V_i$, $V_j$, $V_k$, $V_l$, $V_n$, $V_p$
and $V_q$ (see~\cite{lakshminarayan2014embedded} for a discussion of the specific details, which depend on the prescribed temperature boundary
condition (isothermal or adiabatic material interfaces) and the chosen turbulence model (RANS, LES)). Then, the second-order EBM FIVER 
computes the contributions of the mixed primal cell $(V_m, V_n, V_k, V_j)$ to the vector of numerical diffusive fluxes $\bm{G}$ 
exactly as a standard, second-order, FV method computes the contributions to $\bm{G}$ of any primal cell of a body-fitted mesh.

\begin{remark}Like most if not all other EBMs, IBMs, and GFMs, the second-order EBM FIVER uses the concept of a ghost fluid node and 
populates variables of the fluid state vector at such a node. However, unlike the classical GFM~\cite{fedkiw1999non} and many 
variants, the EBM FIVER populates ghost fluid nodes only in mixed primal cells, and these arise only one graph distance away from 
the boundary of an embedded discrete surface. Hence, this EBM populates only a fraction of the ghost fluid nodes populated by GFM and
related methods.
\end{remark}

Here, a noteworthy update on how to populate variables of the fluid states at ghost fluid nodes of an embedding CFD mesh for the 
purpose of computing the numerical diffusive fluxes is provided. This update is motivated by slender embedded discrete surfaces such 
as those associated with cables which typically lead to mixed primal cell configurations such as those illustrated in~\Cref{fig:GFM}. 
In this figure, the population of $\bm{v}$ and $T$ at the ghost fluid node $V_m$ using information extracted
from the fluid state vectors at the real neighboring nodes $V_i$, $V_j$, $V_k$, $V_l$, $V_n$, $V_p$, and $V_q$ is referred to here as 
a {\it global} approach for populating variables of a fluid state vector at a ghost node. This approach is adopted by many EMBs, IBMs,
and GFMs (for example, see~\cite{tseng2003ghost}). For CFD and FSI applications involving a slender embedded discrete surface, it is 
not appropriate however as it may lead to unphysical computations that promote numerical instability. Indeed, in the aforementioned 
example, the real fluid nodes $\{V_i, V_j, V_k, V_l\}$ and the real fluid nodes $\{V_n, V_p, V_q\}$ lie on two different sides of 
$\Sigma_h$ where, even if the fluid medium is the same, the fluid flow may have totally different states: for example, it could be 
laminar on one side but turbulent on the other, or subsonic on one side but transonic or supersonic on the other, etc. Hence, 
combining in this case components of the fluid state vectors extracted from the union of these two sets of nodes, because they happen 
to be connected to the ghost fluid node $V_m$, in order to populate at this node counterpart variables of the fluid state vector is 
inappropriate. It explains the numerical instabilities that are observed when the global approach for populating $\bm{v}$ and $T$ at 
a ghost fluid node is applied in CFD and FSI applications involving embedded slender structures.

Here, the issue exposed above is addressed by proposing an alternative approach for populating variables of the fluid state vector at
a ghost fluid node, for cable-driven and other FSI problems featuring slender structural subsystems. This approach avoids the 
inappropriate mixing of widely different fluid state vectors without having to identify ``sides'' of an embedded discrete surface. 
It is labeled here as the {\it local} approach -- in contrast with the global approach explained above -- as it consists of populating
at each fluid ghost node $V_m$ connected to one or multiple mixed primal cells one pair of velocity vector $\bm{v}$ and temperature $T$
per mixed cell to which it is connected. For example in the case of the scenario illustrated in~\Cref{fig:GFM}, four different 
velocity-temperature pairs $(\bm{v}, T)$ are populated at $V_m$ as follows:

\begin{itemize}
		\item One pair $(\bm{v}, T)$ computed from information extracted from the fluid state vectors at the real nodes
			$V_i$, $V_j$, and $V_l$, to be used for computing the contributions of the mixed primal cell
			$(V_i, V_j, V_m, V_l)$ to the vector of numerical diffusive fluxes $\bm{G}$.
		\item One pair $(\bm{v}, T)$ computed from information extracted from the fluid state vectors at the real nodes
			$V_j$, $V_k$, and $V_n$, to be used for computing the contributions of the mixed primal cell
			$(V_j, V_k, V_n, V_m)$ to the vector of numerical diffusive fluxes $\bm{G}$.
		\item One pair $(\bm{v}, T)$ computed from information extracted from the fluid state vectors at the real nodes
			$V_l$, and $V_p$, to be used for computing the contributions of the mixed primal cell
			$(V_l, V_m, V_p, V_o)$ to the vector of numerical diffusive fluxes $\bm{G}$.
		\item One pair $(\bm{v}, T)$ computed from information extracted from the fluid state vectors at the real nodes
			$V_n$, $V_q$, and $V_p$, to be used for computing the contributions of the mixed primal cell
			$(V_m, V_n, V_q, V_p)$ to the vector of numerical diffusive fluxes $\bm{G}$.
\end{itemize}

\subsection{Loads computation}
\label{sec:load_comp}

\begin{figure}[!h]
\centering
\includegraphics[width=0.7\linewidth]{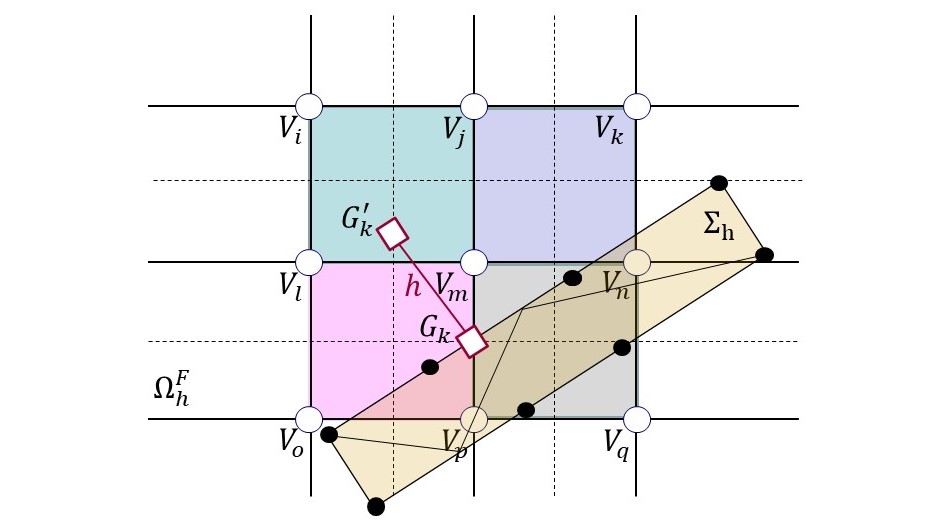}
	\caption{Shifting by a distance $h$ of a Gauss point $G_k$ used for evaluating the flow-induced forces on the material interface $\Sigma_h$ (two-dimensional case).}
\label{fig:GP}
\end{figure}

A critical component of the master-slave kinematic approach for cable-driven FSI described in \Cref{sec:alternative approach}
is the computation of the force vector $\bm{f}^n_{S^j_i}$ at each slave node $S_i^j$ as specified in (\ref{eq:fs}), from which
the force and moments vectors at the master point $M_i$, $\bm{f}^n_{M_i}$ and $\bm{m}^n_{M_i}$ (\ref{eq:load_M}), respectively, 
and the generalized loads $\bm{f}_{N_j}^n$ and $\bm{m}_{N_j}^n$ (\ref{eq:load_N}) can be deduced. This force vector is best
evaluated by performing the integration in (\ref{eq:fs}) directly on the embedded, discrete surface $\Sigma_h$~\cite{lakshminarayan2014embedded}
and approximating it using a Gauss quadrature rule. Such an approximation can be written as

\begin{equation} 
\bm{f}^n_{S^j_i} = \int_{\Sigma_h^n} \Big( -p^n \bm{n}^n + \bm{\tau}^n \bm{n}^n\Big)\phi_{S^j_i}\,d\Sigma_h  \approx \sum_{k=1}^{n_{G}} \varpi_k\Big( -p^n_{G_k} \bm{n}_{G_k}^n 
+ \bm{\tau}^n_{G_k}\bm{n}^n_{G_k}\Big)[\phi_{S^j_i}]_{G_k}
\label{eq:GP} 
\end{equation}
where $\varpi_k$ denotes the $k$-th weight of the quadrature rule, and the subscript $G_k$ designates the evaluation of a quantity at the $k$-th quadrature point $G_k$ associated with the
weight $\varpi_k$. Each Gauss quadrature point $G_k \in \Sigma_h$ can be located in some primal cell $\mathcal C_l$ of the embedding CFD mesh. Hence, $p^n_{G_k}$ and $\bm{\tau}^n_{G_k}$ are typically 
computed by interpolation of fluid state variables attached to the nodes of $\mathcal C_l$. However, $\mathcal C_l$ is typically a mixed primal cell connected to ghost fluid nodes -- that is, nodes of 
the embedding CFD mesh that are occluded by $\Sigma_h$ (for example, see~\Cref{fig:GP}) -- and the populated the fluid state variables at these ghost nodes are typically tainted by nonphysical 
values due to the underlying extrapolations (see~\cite{huang2018family}). Consequently, the approximation~(\ref{eq:GP}) is typically fraught with spurious oscillations.
For this reason, it was proposed in~\cite{huang2018family} to shift each Gauss point $G_k$ used in the approximation~(\ref{eq:GP})
in the direction of the outward normal $\bm{n}_{G_k}$ to the material interface $\Sigma_h$, up to the point $G_k^{\prime}$ defined by (see~\Cref{fig:GP})
\begin{equation*}
	\bm{x}_{G_k^{\prime}} = \bm{x}_{G_k}  + h \bm{n}_{G_k}
\end{equation*}
where $h$ is the characteristic mesh size of the fluid primal cell $\mathcal C_l$. 

Indeed, the shifted Gauss point $G_k^{\prime}$ will be generally located in this case in a primal cell without
any ghost fluid node, the interpolations of $p^n_{G_k^{\prime}}$ and $\bm{\tau}^n_{G_k^{\prime}}$ will not be tainted by any nonphysical data, and unlike the approximation (\ref{eq:GP}), the 
quadrature approximation
\begin{equation} 
	\bm{f}^n_{S^j_i} = \int_{\Sigma_h^n} \Big( -p^n \bm{n}^n + \bm{\tau}^n \bm{n}^n\Big)\phi_{S^j_i}\,d\Sigma_h  \approx \sum_{k=1}^{n_{G}} \varpi_k\Big( -p^n_{G_k^{\prime}} \bm{n}_{G_k^{\prime}}^n 
	+ \bm{\tau}^n_{G_k^{\prime}}\bm{n}^n_{G_k^{\prime}}\Big)[\phi_{S^j_i}]_{G_k^{\prime}}
	\label{eq:GPP} 
\end{equation}
will not exhibit any spurious oscillation. 

Furthermore, from $\displaystyle{\frac{\partial p_{G_k}}{\partial \bm{n}}} \approx 0$, due to the conservation of momentum in the normal direction at the 
material interface, and the Taylor expansion of the pressure around the point $G_k$, it follows that
\begin{equation*}
	        p_{G_k^{\prime}} = p_{G_k} + O(h^2)
\end{equation*}
The above result implies that the contribution of the pressure term to the approximation~(\ref{eq:GPP}) is a third-order approximation of its counterpart for~(\ref{eq:GP}). However,
a similar Taylor expansion shows that the contribution of the shear stress traction to the approximation~(\ref{eq:GPP}) is a second-order approximation of its counterpart for the 
approximation~(\ref{eq:GP}). Hence, the loads evaluated using the approximation~(\ref{eq:GPP}) are a second-order approximation of the loads evaluated using~(\ref{eq:GP}).

\subsection{Adaptive mesh refinement}
\label{sec:AMR}

When the embedded discrete surface undergoes large motions and/or deformations, maintaining reasonable boundary layer 
resolution around $\Sigma_h$ requires special effort in the Eulerian computational framework for FSI. At high Reynolds numbers,
it requires for this purpose 
AMR~\cite{berger1989local, maubach1995local, stevenson2008completion, vanella2014adaptive, roma1999adaptive}. All FIVER methods,
including the second-order FIVER method described above are equipped with a local edge refinement and coarsening algorithm based 
on the newest vertex bisection~(NVB) method~\cite{maubach1995local, stevenson2008completion, borker2019mesh}, which enables the 
boundary layer and flow features to be efficiently tracked using a wall distance estimator and a Hessian-based error indicator, 
respectively. For a cable subsystem however, fully resolving its boundary layer is generally unaffordable, especially when the cable 
has a large length-to-diameter ratio. 

To address the issue raised above, a new criterion for marking edges for refinement is proposed here. This criterion constitutes a 
lightweight alternative to the wall distance criterion originally proposed in~\cite{borker2019mesh}. Specifically, wherever an edge of 
the embedding fluid mesh $\Omega_h^F$ is intersected twice by the embedded discrete surface $\Sigma_h$ -- which indicates that the 
computational fluid domain is underresolved in this region -- the edge is selected for refinement and subsequently bisected. This 
criterion leads to an affordable mesh. Most importantly, it enables capturing the cable-driven FSI effects. This new edge refinement 
criterion, which is referred to in the remainder of this paper as the \emph{doubly-intersected} edge criterion, is equally applicable
and equally important for FSI problems where the structural system contains small-scale subsystems that are typically unresolved by 
practical CFD meshes. It is illustrated in \Cref{sec:parachute}, where its effectiveness is also demonstrated. 

\section{Applications}
\label{sec:apps}

The dressing approach and the proposed master-slave kinematic approach for cable-driven FSI are implemented in the nonlinear, fluid-structure
simulation platform AERO Suite (for example, see~\cite{geuzaine2003aeroelastic}). This software suite includes, among other modules: the versatile, nonlinear, structural
analyzer AERO-S; and the comprehensive, compressible, flow solver AERO-F. It has been validated for numerous wind tunnel, flight test, and other fluid-structure configurations 
pertaining to various underwater systems, high performance cars, and aircraft. It supports both the ALE and Eulerian computational frameworks for FSI.

Here, the Eulerian computational framework of AERO Suite and its EBM FIVER are applied to the simulation of two different FSI problems in order to illustrate, verify, and validate the
proposed master-slave kinematic approach for cable-driven FSI computations:
\begin{itemize}
	\item An airborne refueling model problem, where the cable subsystem consists of a single flexible hose that undergoes large motions, and FSI is captured using implicit-implicit
		fluid-structure computations. In this case, the dressing approach is used to verify the accuracy of the proposed master-slave kinematic approach.
	\item A dynamic, supersonic parachute inflation problem where the cable subsystem consists of 80 suspension lines, the canopy and suspension lines undergo large motions and deformations,
		and FSI is captured using implicit-explicit fluid-structure computations due to the massive self-contact experienced by the canopy during the inflation process. The dressing
		approach is not suitable for this problem for the reason discussed in~\Cref{sec:dressing approach}. The proposed master-slave kinematic approach
		is validated in this case using real flight data.
\end{itemize}

\subsection{Airborne refueling model problem}

\begin{figure}[!h]
  \centering
  \includegraphics[width=0.6\textwidth]{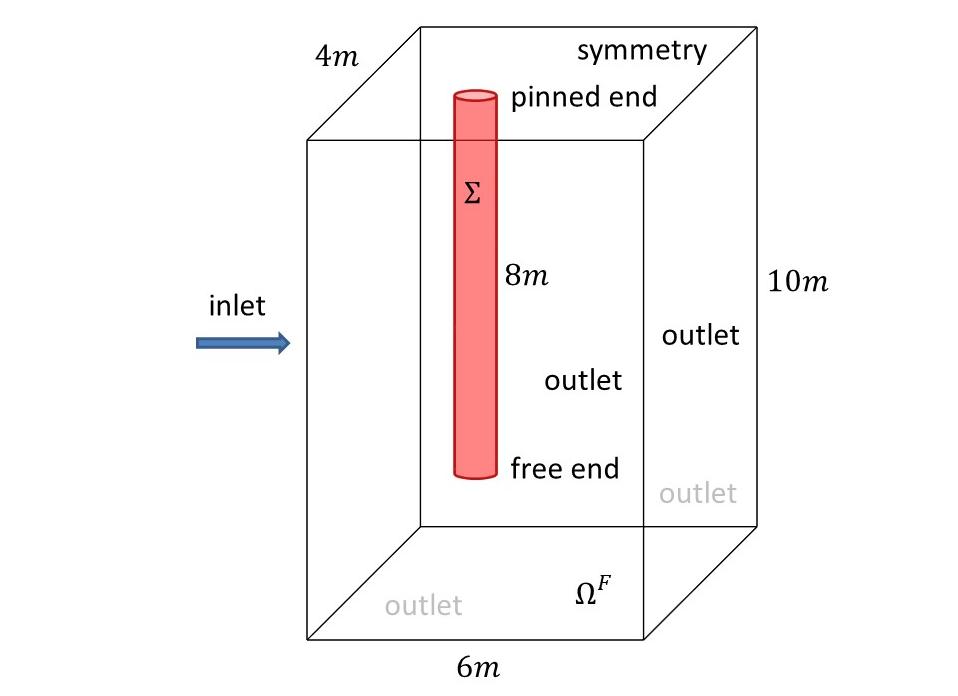}
  \caption{Airborne refueling model problem: hose and embedding computational fluid domain.}
  \label{fig:ARS_schematics}
\end{figure}

First, an airborne refueling model problem~\cite{ro2010modeling, zhu2007modeling, styuart2011numerical} consisting of predicting the dynamic, aeroelastic response of a flexible hose that trails from 
an aircraft tanker to the unsteadiness of the flow surrounding it is considered. This problem, which is graphically depicted in~\Cref{fig:ARS_schematics}, is designed to compare the performance of the 
proposed master-slave kinematic approach for modeling cable-driven FSI to that of the dressing approach. The flexible hose has a length of $8$~m, and a length-to-diameter ratio of $L/D = 119.4$. Its 
linear elastic material and other geometrical properties are given in the top part of~\Cref{tab:airborne_refueling_system}. The high-speed inflow conditions are given in the bottom part of this figure: 
they correspond to an altitude of $8$~km above sea level. Air viscosity is modeled using Sutherland's viscosity law with the constant $\mu_0 = 1.458 \times 10^{-6}$ ~kg m$^{-1}$s$^{-1}$ and reference 
temperature $T_0 = 110.6$~K. The Reynolds number based on the hose diameter is approximately $3.9 \times 10^{5}$. Hence, the flow is assumed to have already transitioned to the turbulent regime, which 
is modeled here using the Spalart-Allmaras turbulence model~\cite{spalart1992one}.

\begin{table}[!htbp]
\centering
\begin{tabular}{@{}lll@{}}
\toprule
 Parameter & Description & Value \\ 
\midrule
 $L$       & Length               & 8 m              \\ 
 $m_L$     & Mass per unit length & 0.38 kg m$^{-1}$ \\ 
 $D$       & Diameter             & 0.067 m          \\
 $E$       & Young's modulus      & 17 MPa           \\
 $\nu$     & Poisson's ratio      & 0.42             \\
\midrule 
 $H$             & Altitude                  & 8 km             \\
 $\rho_{\infty}$ & Free-stream density       & 0.58 kg m$^{-3}$ \\
 $p_{\infty}$    & Free-stream pressure      & 40000 Pa         \\
 $T_{\infty}$    & Free-stream temperature   & 240 K            \\
 $M_{\infty}$   & Free-stream Mach number   & 0.5              \\
\bottomrule
\end{tabular}
	\caption{Geometrical and material properties of the hose~\cite{ro2010modeling, styuart2011numerical} (top); and inflow conditions (bottom).}
\label{tab:airborne_refueling_system}
\end{table}
\vglue 0.2truein

The hose is discretized by 100 geometrically nonlinear beam elements, and its surface is represented by a uniform discretization characterized by 1,200 triangular elements and a uniform, hexagonal,
discrete cross section. It is pinned at one end -- that is, all 3 translational DOFs of its FE model are fixed at one end -- and unrestrained at the other. Hence, it is anticipated that this flexible 
hose will undergo large motions in the presence of a high-speed airstream. 

The embedding computational fluid domain $\Omega^F$ (see \Cref{fig:ARS_schematics}) is chosen to be a 6~m $\times$ 4~m $\times$ 10~m cube. It is initially discretized by 67,686 tetrahedra. During the 
unsteady FSI simulation, this embedding CFD mesh is adaptively refined or coarsened, as needed, using: a distance-based criterion to track and resolve the geometry of the hose as well as the boundary 
layer; a velocity magnitude Hessian  criterion to capture the shedded vortices; and an algorithm based on the newest vertex bisection algorithm~\cite{stevenson2008completion, borker2019mesh} to adapt 
the edges of the CFD mesh. Consequently, the characteristic size of the embedding CFD mesh near the hose is about 4 $\times$ 10$^{-3}$~m -- that is, roughly 1/20$^\textrm{th}$ of its diameter. Away from 
the boundary layer, the minimum edge length of the embedding CFD mesh is set to 1.5 $\times$ 10$^{-2}$~m.

Time-discretization of the structural subproblem is performed using the implicit midpoint rule, that of the fluid subproblem is performed using the 3-point implicit Backward Difference Method (BDF),
and both discretizations are coupled using the second-order, time-accurate, implicit-implicit fluid-structure staggered solution procedure presented in~\cite{farhat2010robust}. Two FSI simulations
are performed using the coupling time-step of $\Delta t_{F/S} = 2\times 10^{-6}$~s in order to compute the aeroelastic response of the hose in the time-interval $[0, 0.04]~s$: one using the dressing 
approach; and another using the proposed master-slave kinematic approach.

\begin{figure}[!h]
  \centering
  \includegraphics[width=0.24\textwidth]{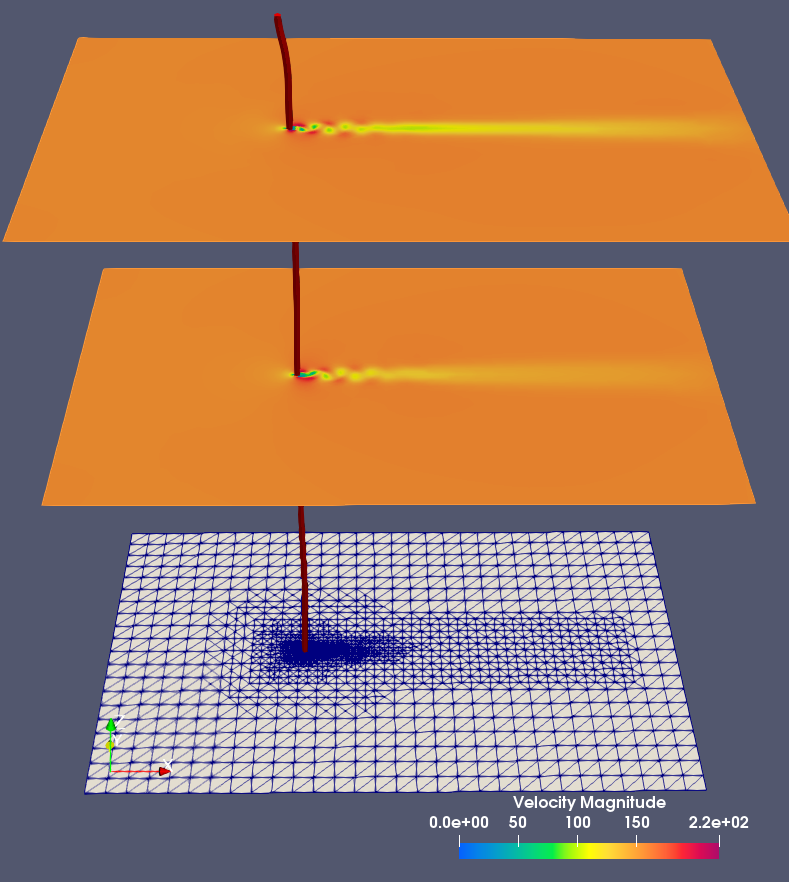}
  \includegraphics[width=0.24\textwidth]{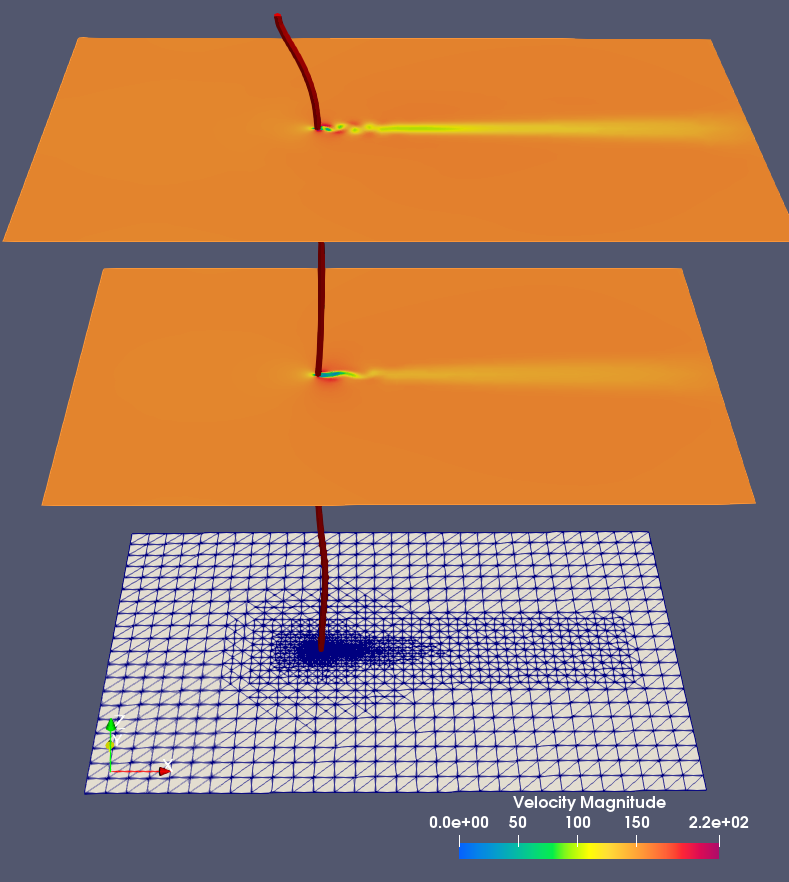}
  \includegraphics[width=0.24\textwidth]{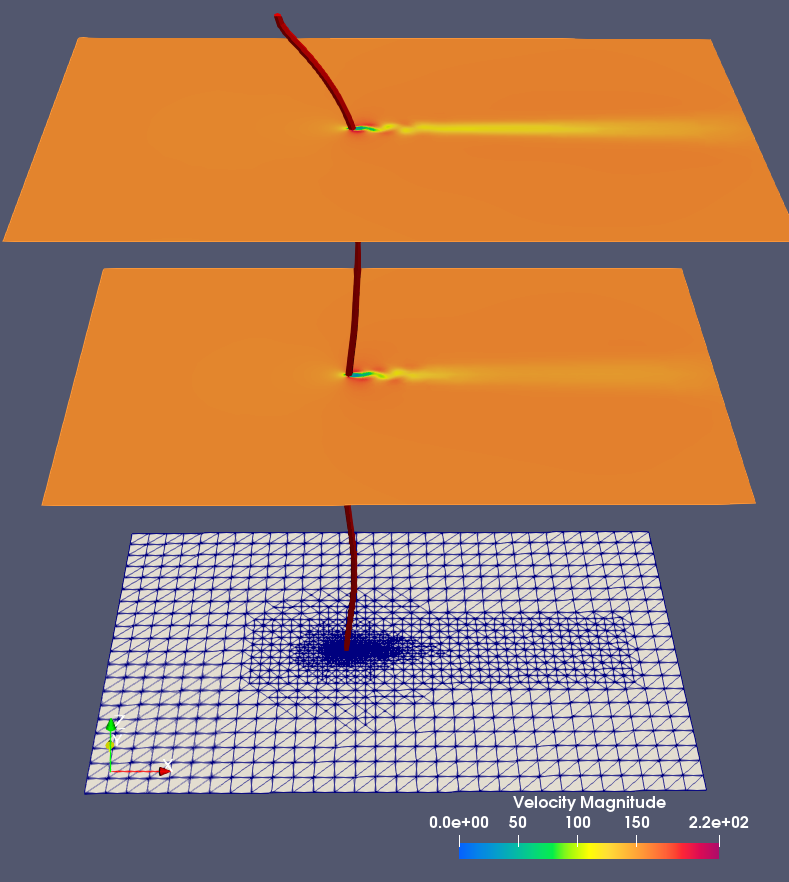}
  \includegraphics[width=0.24\textwidth]{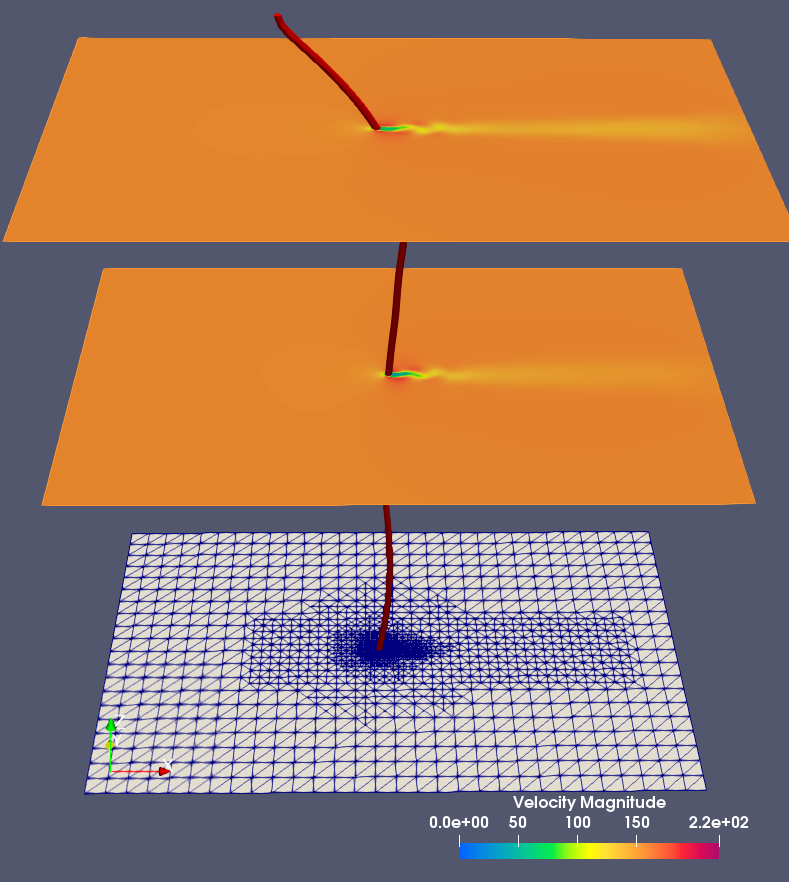}\\
\vspace{2pt}
  \includegraphics[width=0.24\textwidth]{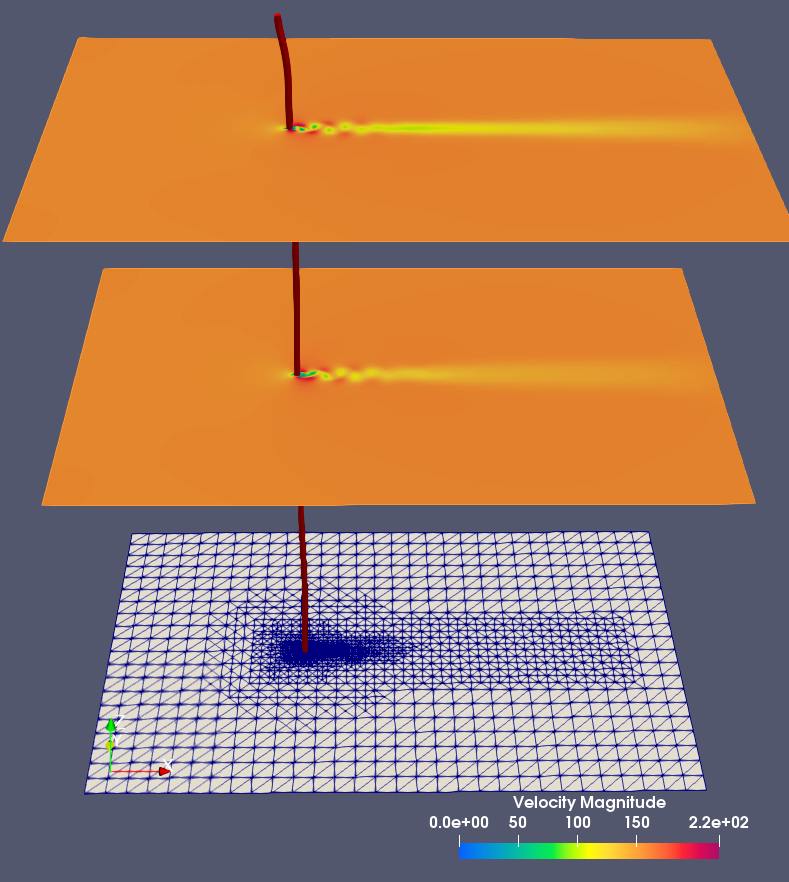}
  \includegraphics[width=0.24\textwidth]{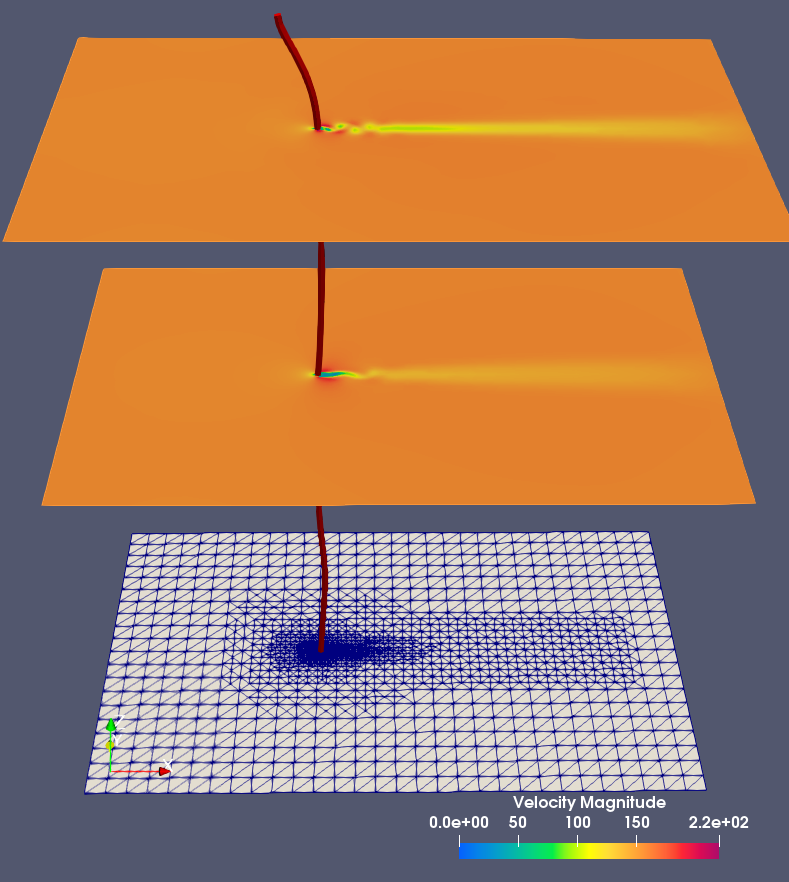}
  \includegraphics[width=0.24\textwidth]{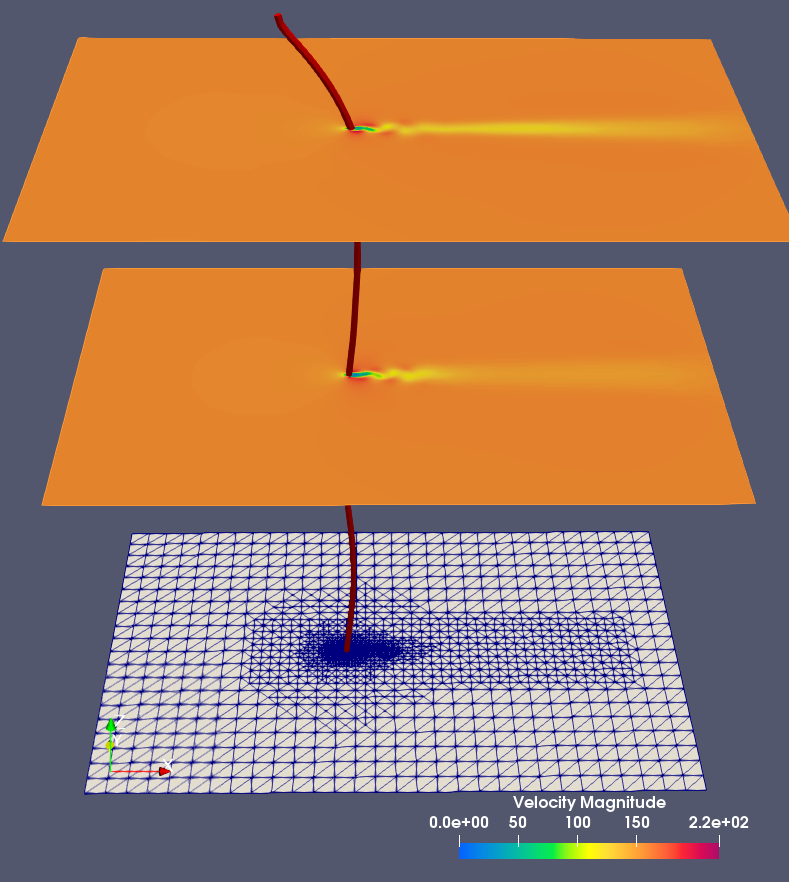}
  \includegraphics[width=0.24\textwidth]{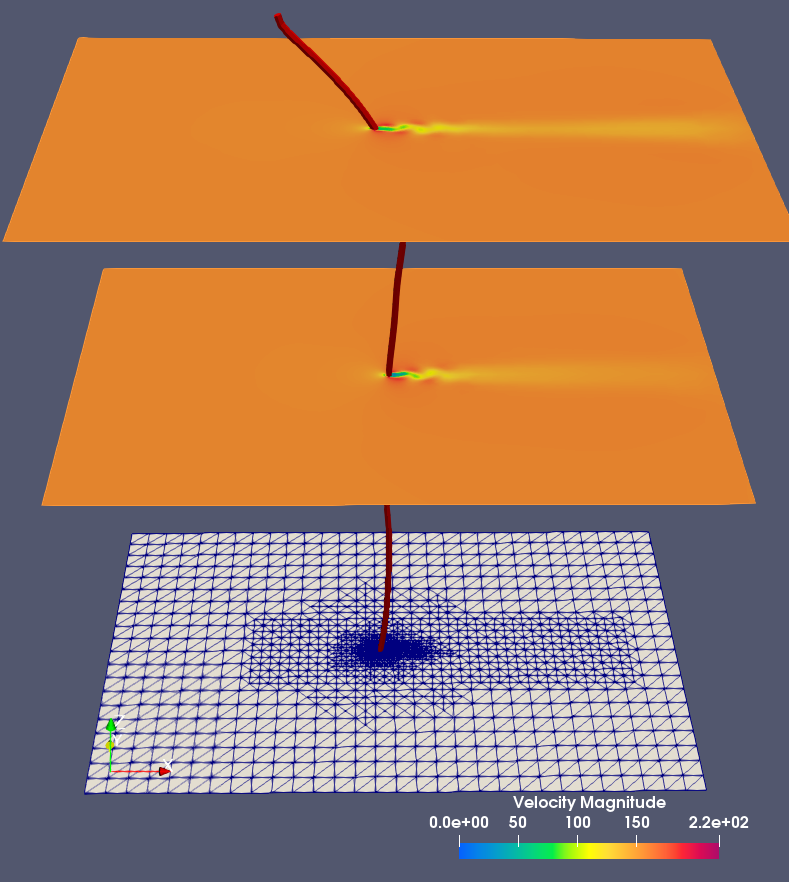}
  \caption{Structural motion and velocity magnitude field in cross sections of $\Omega^F$ computed using: the dressing approach (top); and the master-slave kinematic approach~(bottom) -- Solution
	snapshots and adapted embedding CFD mesh at to $t = 0.01$~s, $t = 0.02$~s, $t = 0.03$~s, and $t = 0.04$~s (left to right): cross section of $\Omega^F$ at the pinned end of the hose (top); cross 
	section of $\Omega^F$ at the middle section of the hose (middle); and cross section of $\Omega^F$ at the free end of the hose (bottom).}
\label{fig:ARS_velocity}
\end{figure}

\Cref{fig:ARS_velocity} illustrates the time-evolution of the adapted embedding CFD mesh at the cross section of $\Omega^F$ corresponding to the free end of the hose, and the velocity magnitude of the 
fluid flow and displacement of the hose at two different cross sections of the computational fluid domain: that corresponding to the middle section of the hose; and that corresponding to the free end of 
the hose where the structural motion is the largest. It reveals that in response to the high-speed flow, the hose drifts while interacting with the trailing vortices. This figure also shows that AMR 
effectively tracks the boundary layer and flow features. 

\begin{figure}[!h]
  \centering
  \includegraphics[width=0.49\textwidth]{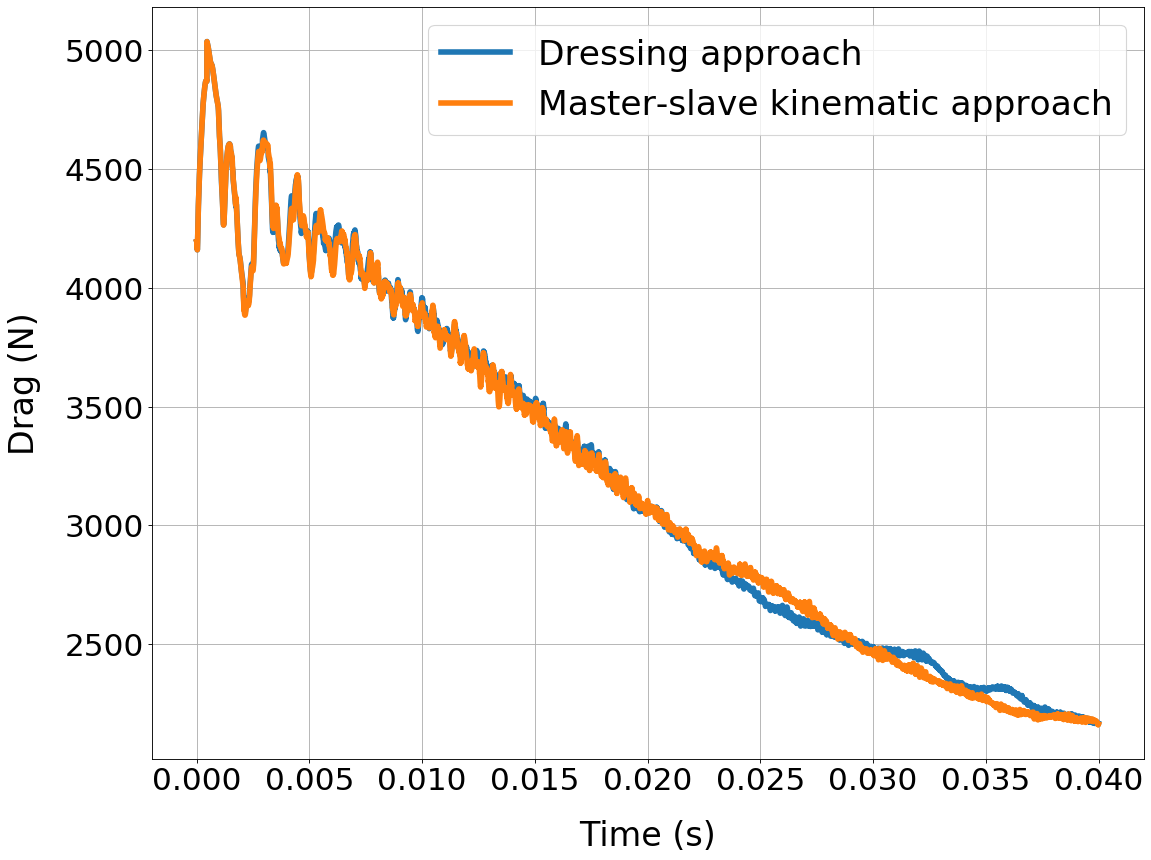}
  \includegraphics[width=0.49\textwidth]{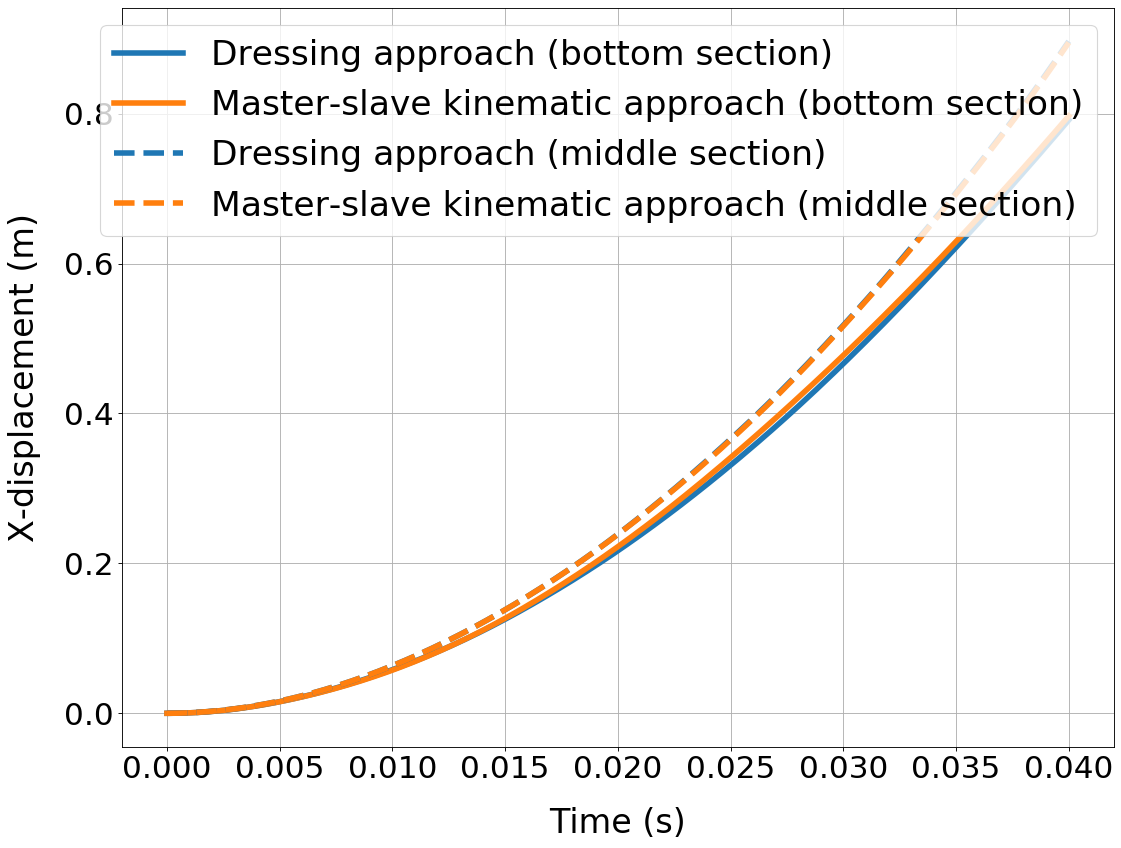}
	\caption{Time-histories of the drag (left) and lateral displacements of the hose at different sections (right) computed using: the dressing approach (blue); and the alternative master-slave 
	kinematic approach (orange).}
\label{fig:ARS_drag_disp}
\end{figure}

More importantly, ~\Cref{fig:ARS_velocity} shows that both the dressing and master-slave kinematic approaches deliver almost the same fluid and structural responses. This conclusion is supported 
by ~\Cref{fig:ARS_drag_disp} which shows that the FSI simulation equipped with the master-slave kinematic approach produces the same time-histories of the lateral displacement of the hose and the drag 
force acting on it as its counterpart equipped with the dressing approach.

\subsection{Dynamic supersonic parachute inflation problem}
\label{sec:parachute}

Next, the FSI simulation of the inflation dynamics of a Disk-Gap-Band (DGB) parachute system in the low-density, low-pressure, supersonic Martian atmosphere is 
considered~\cite{cruz2014reconstruction}. While such a simulation is crucial to the understanding of the effects of the FSI driven by the suspension line
subsystem on the performance of the main parachute during the deceleration process, its main purpose here is to {\it validate} the proposed master-slave approach for cable-driven FSI using flight data
from the landing on Mars of NASA's rover Curiosity.

\begin{figure}[!h]
  \centering
	\includegraphics[width=0.7\textwidth]{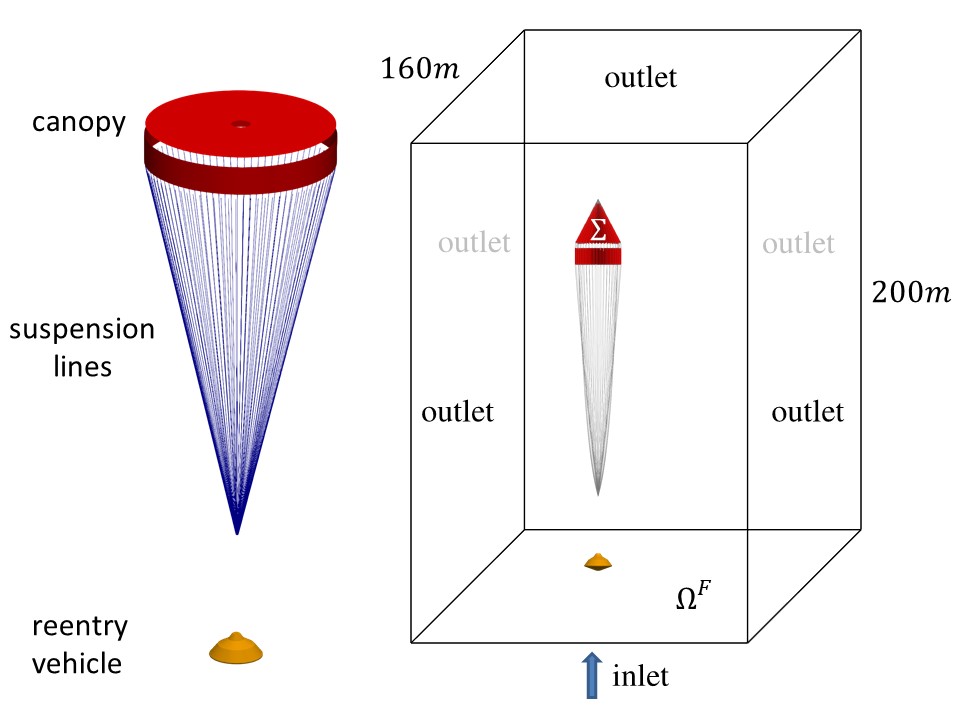}
	\caption{Dynamic supersonic parachute inflation problem: system configuration (left); and embedding computational 
	fluid domain as well as embedded initial folded configuration (right).}
	\label{fig:PIDCFD}
\end{figure}

To this end, the DGB parachute system that successfully deployed in 2012 for the Mars landing of Curiosity is considered (see~\Cref{fig:PIDCFD}-left). This aerodynamic decelerator system
consists of three main components~\cite{cruz2014reconstruction}:
\begin{itemize}
\item The canopy, which is made of F-111 nylon.
\item The suspension lines, which are made of Technora T221 braided cords.
\item And the reentry vehicle.
\end{itemize}
Its exact geometric and material properties are listed in~\Cref{tab: parachute}.  

The simulation discussed herein starts from the line stretch stage where the suspension line subsystem is deployed but the canopy is folded (see~\Cref{fig:PIDCFD}-right), and the entire system is prestressed by the folding pattern. The incoming supersonic flow is at 
the state defined by $M_{\infty} = 1.8$, $\rho_{\infty} = 0.0067$~kg~m$^{-3}$, and $p_{\infty} = 260$~Pa. 

\begin{table}
\centering
\begin{tabular}{@{}llll@{}}
\toprule
Component &  Parameter & Description & Value\\ 
\midrule
Canopy & $D$      & Diameter        & 15.447 m                   \\
       & $t$      & Thickness       & 7.607 $\times$ 10$^{-5}$ m \\
       & $E$      & Young's modulus & $0.945$ GPa                \\
       & $\nu$    & Poisson's ratio & 0.4                        \\
       & $\rho^C$ & Density         & 1154.25 kg m$^{-3}$        \\
       & $\alpha$ & Porosity        & 0.08                       \\ 
\midrule
Suspension lines & $L$      & Length          & 36.56 m                    \\
                 & $D$      & Diameter        & 3.175 $\times$ 10$^{-3}$ m \\
                 & $E$      & Young's modulus & 29.5 GPa                   \\
		 & $\rho^{SL}$ & Density         & 1154.25 kg m$^{-3}$        \\
\bottomrule
\end{tabular}
\caption{True geometrical and material properties of a DGB parachute system~\cite{lin2010flexible, cruz2014reconstruction}.}
\label{tab: parachute}
\end{table}

Since the Martian atmosphere is mainly composed of carbon dioxide, the viscosity of this gas is modeled using Sutherland's viscosity law with the constant $\mu_0 = 1.57\times10^{-6}$~kg m$^{-1}$s$^{-1}$ 
and the reference temperature $T_0 = 240$~K. The Reynolds number based on the canopy diameter is $4.06 \times 10^6$. Hence, the flow is assumed to have transitioned to the turbulent regime, which is 
modeled here using Vreman's eddy viscosity subgrid-scale model for turbulent shear flow~\cite{vreman2004eddy} with model constant $C_s = 0.07$. 

The canopy of the DGB parachute consists of band and disk gores. Here, these are discretized by 279,025 geometrically nonlinear thin shell ANDES elements~\cite{militello1991first} 
(although the membrane stiffness of the fabric is significantly larger than its bending stiffness, both stiffnesses are considered here). The suspension line subsystem contains 80 lines, each 
of which is discretized by 500 geometrically nonlinear beam elements. As for the reentry vehicle, it is modeled as a fixed rigid body that is embedded, together with the entire aerodynamic decelerator 
system, in the embedding computational fluid domain (see~\Cref{fig:PIDCFD}).

The aforementioned computational fluid domain is a box of size 200 m $\times$ 160 m $\times$ 160 m. It is initially discretized by a mesh composed of Kuhn 
simplices~\cite{stevenson2008completion, borker2019mesh}. Specifically, this initial tetrahedral mesh contains 2,778,867 nodes and 16,308,672 tetrahedra. During the FSI simulation reported below,
AMR~\cite{mitchell1988unified, stevenson2008completion, borker2019mesh} is applied to track and resolve, as in the the case of the airborne refueling model problem, the boundary layer and flow features. 
The doubly-intersected edge criterion mentioned in~\Cref{sec:AMR} is applied to each suspension line to resolve the flow around it. The specified characteristic mesh sizes near the reentry vehicle, 
suspension lines, and canopy are $2.5$~cm, $3$~mm, and $5$~cm, respectively. The specified characteristic mesh size in the wake and near the shock is $10$~cm.  

Since the canopy is made of F-111 nylon with an $8\%$ void fraction, its permeability is modeled using the homogenized porous wall model proposed in~\cite{huang2019homogenized}, which is
based on the earlier model described in~\cite{huang2018simulation}. Due to the massive self-contact of the parachute canopy during its
dynamic inflation, the explicit central difference time-integration scheme 
is applied for advancing in time the semi-discrete state of the structural subsystem. Consequently, only the master-slave kinematic approach described in~\Cref{sec:alternative approach} is applied here for modeling 
the suspension line-driven FSI. For this purpose, each suspension line is assumed to have a circular cross section with a 
diameter $D = 3.2$ ~mm. Its surface is represented by a uniform discretization
characterized by 6,012 triangular elements and a uniform, hexagonal, discrete cross section.

First, a quasi-steady state of the flow past the folded parachute configuration shown in~\Cref{fig:PIDCFD}-right is computed assuming that this configuration is rigid and fixed. Using this CFD solution and the aforementioned
prestressed state of the structural model of the parachute system as initial fluid and structural conditions, respectively, the FSI simulation of the inflation dynamics of the DGB parachute is performed in the time-interval 
$[0, 0.8]$~s. The length of this time-interval is such that it covers the inflation process as well as a few breathing cycles of the DGB parachute system. As stated above, the explicit central difference time-integrator
is applied to advancing in time the semi-discrete structural subsystem. On the other hand, the implicit, 3-point BDF scheme is applied 
to time-integrate the semi-discrete fluid state. For this reason, the fluid and structural discretizations are coupled for this
simulation using the stability-preserving, second-order, time-accurate, implicit-explicit fluid-structure staggered solution 
procedure presented in~\cite{farhat2010robust}. The fluid-structure coupling time-step is set to $\Delta t_{F/S} = 10^{-5}$~s.

\begin{figure}[ht]
  \centering
  \includegraphics[width=0.32\textwidth]{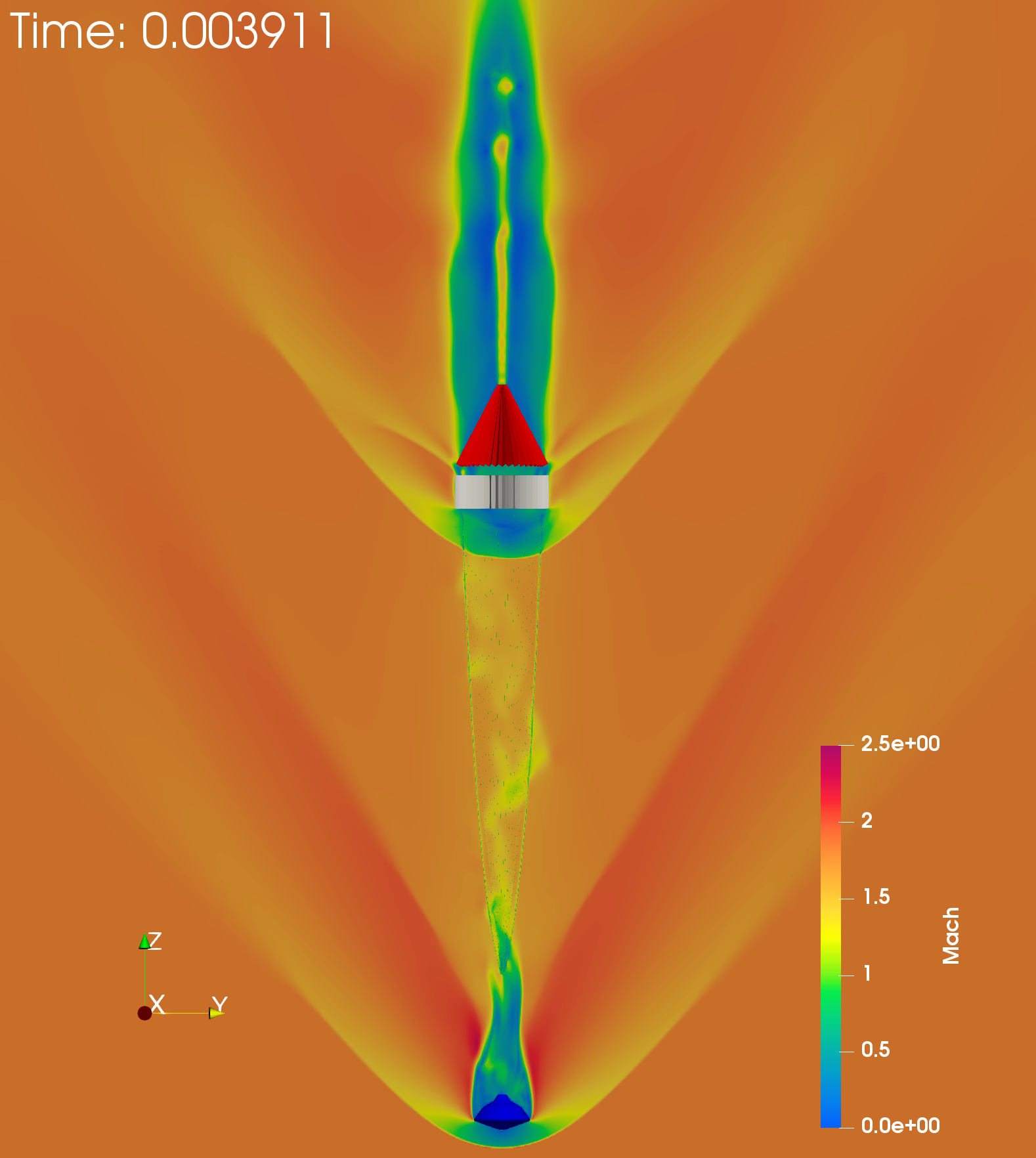}
  \includegraphics[width=0.32\textwidth]{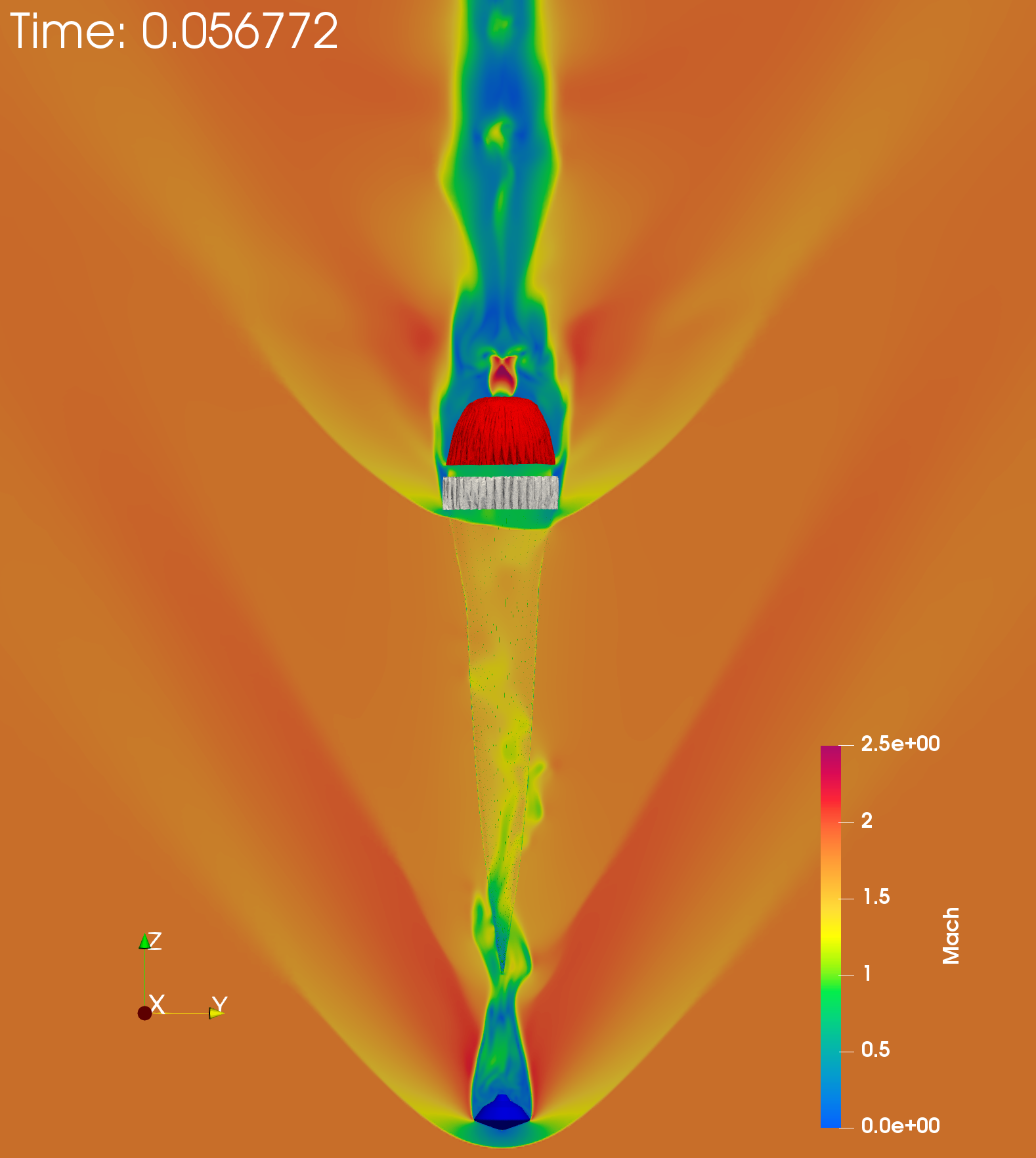}
  \includegraphics[width=0.32\textwidth]{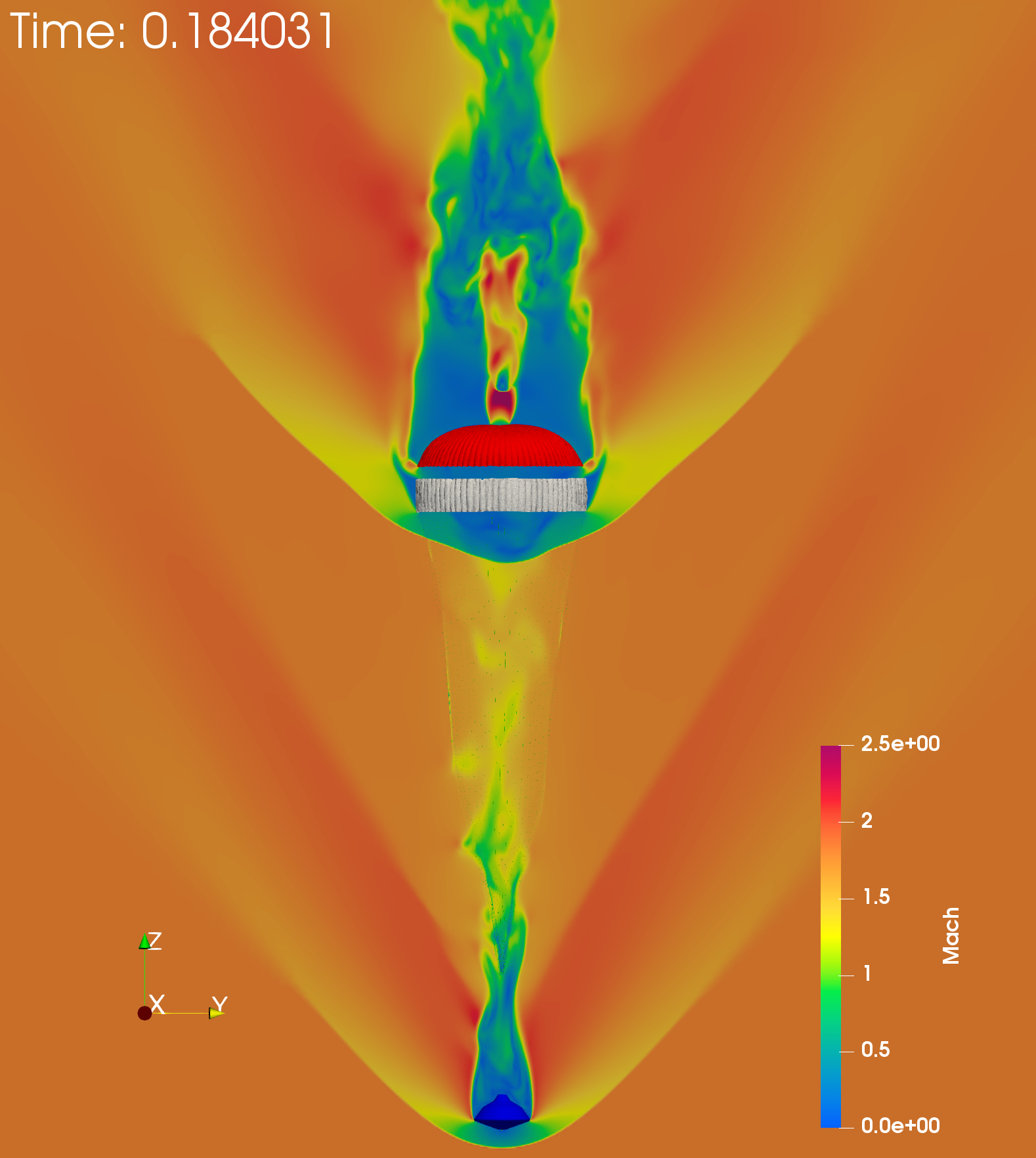}\\
\vspace{2pt}
  \includegraphics[width=0.32\textwidth]{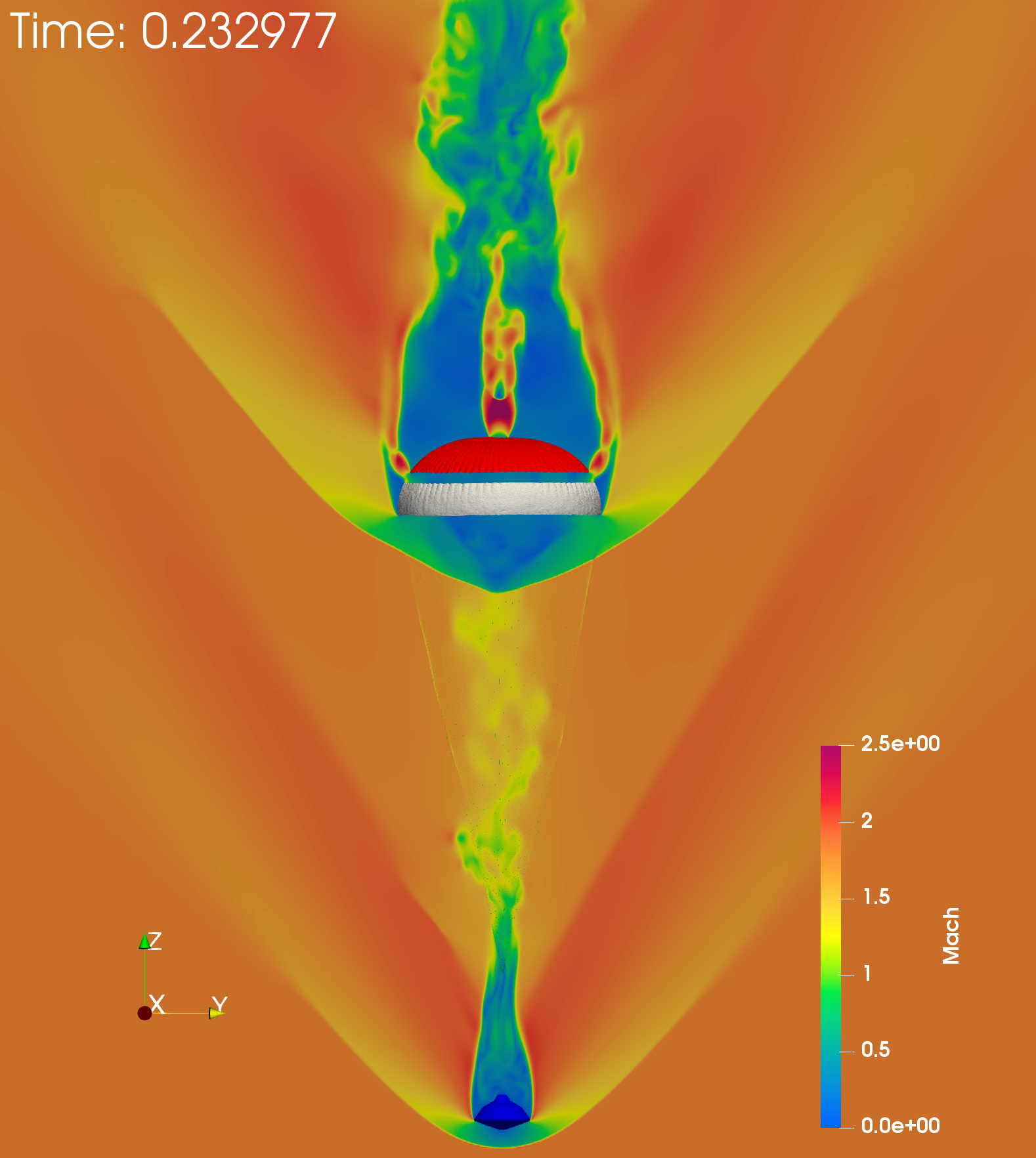}
  \includegraphics[width=0.32\textwidth]{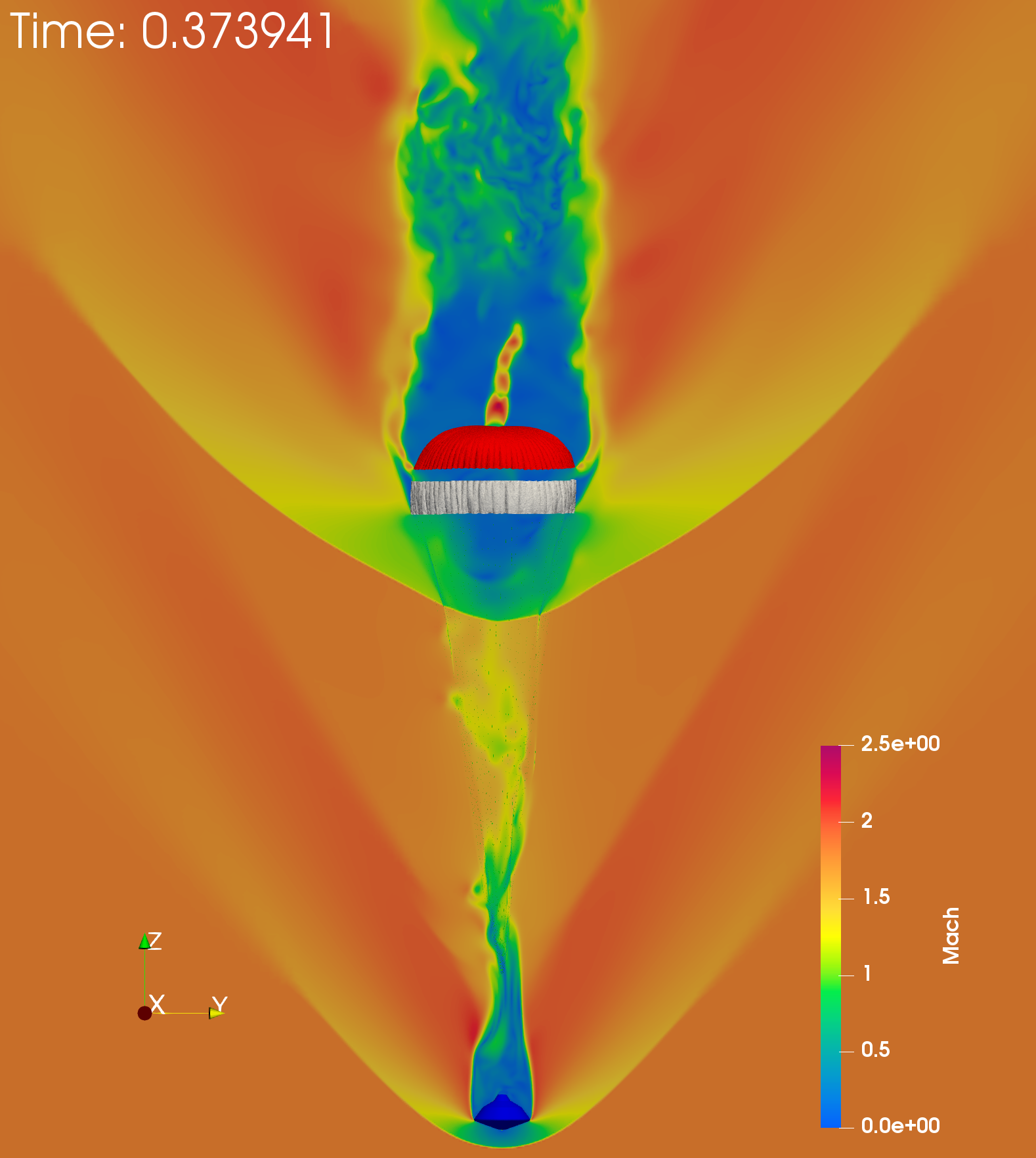}
  \includegraphics[width=0.32\textwidth]{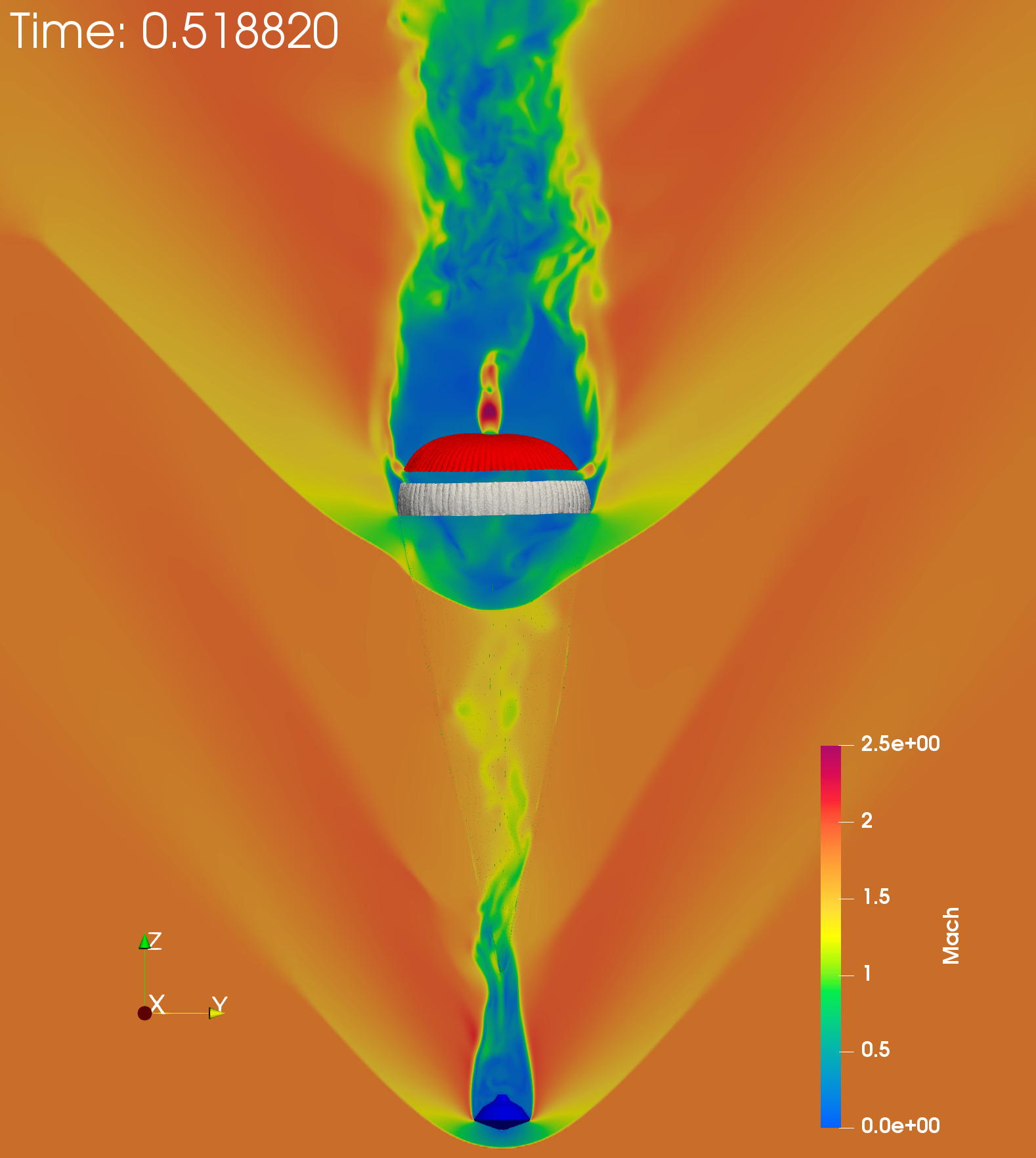}
  \caption{Time-evolutions of the deployment of the parachute DGB system and the associated flow Mach number field.}
  \label{fig:pid_mach}
\end{figure}

Figure~\ref{fig:pid_mach} graphically depicts the time-evolutions of the dynamic inflation of the DGB parachute and the flow Mach 
number around it. It shows that disturbed by the wake generated by the reentry vehicle, the bow shock in the front of the canopy 
vibrates along with the breathing cycles of the canopy. It also reveals that a jet-like flow is ejected through the canopy vent and 
gaps between the band and disk gores, and interacts with the turbulent wake behind the canopy. The parachute is fully inflated at 
approximately $t = 0.23$~s: after this time, it starts the breathing cycles expected from a violent, high-speed, dynamic, inflation 
process.

Although the suspension lines are in this case very slender, they increase the geometric blockage in front of the canopy, 
which slightly reduces the flow speed. Moreover, the contributions of these suspension lines to FSI can be observed 
in~\Cref{fig:pid_mach}-e to generate shock waves that alter the wake behind the reentry vehicle, and even disrupt the bow-shock. 
These observations are in perfect agreement with those documented in~\cite{sengupta2009supersonic} for sub-scale parachute inflation
experiments conducted in wind tunnels.

A second FSI simulation is performed, but without the contribution of the suspension line subsystem to FSI. In other words,
the dynamics of the suspension line subsystem are accounted for in the overall FE structural model, but the effects of this subsystem
on the fluid flow and associated FSI are ignored.

\begin{figure}[ht]
  \centering
  \includegraphics[width=0.8\textwidth]{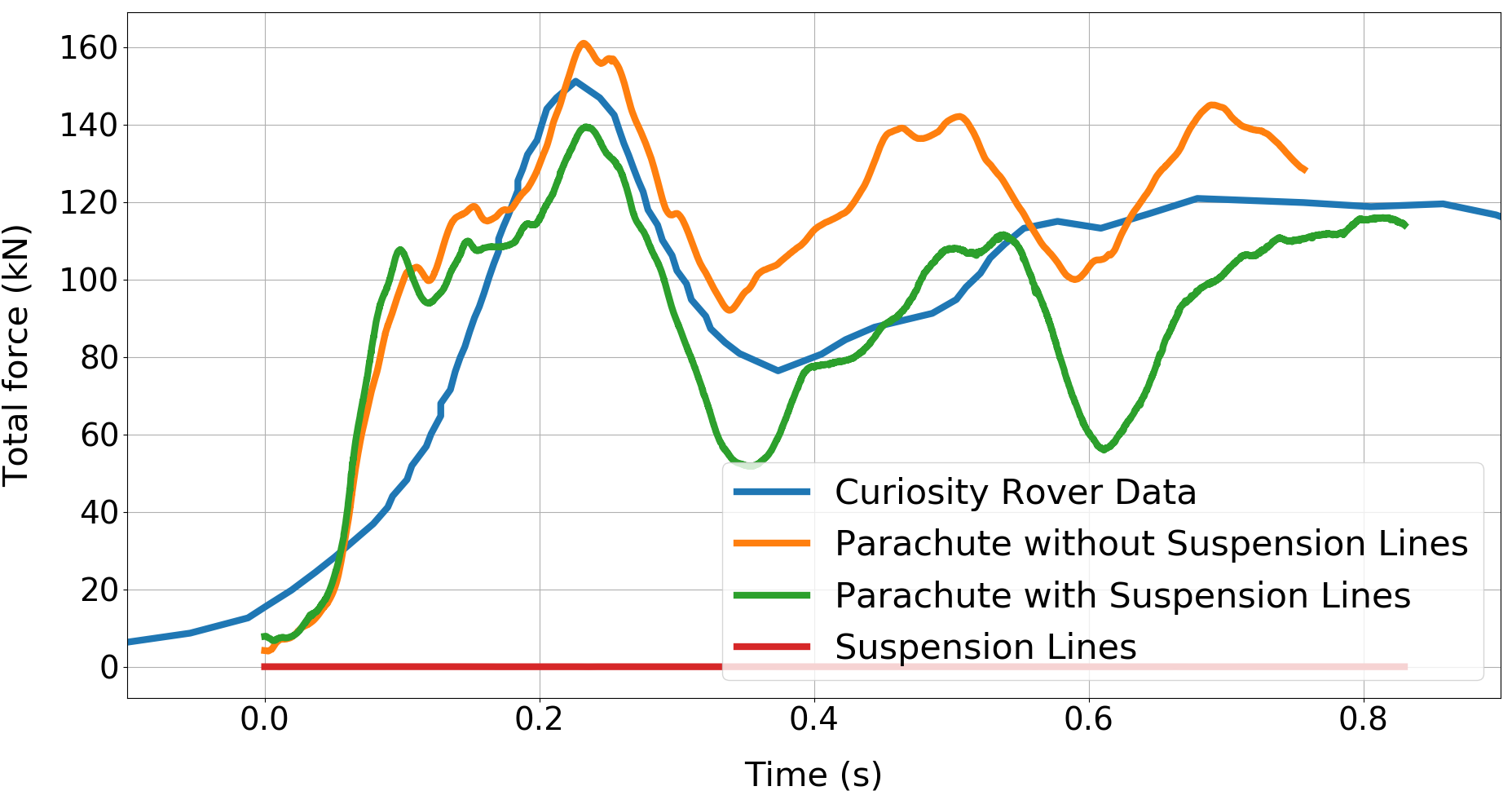}
  \caption{Time-histories of the total drag generated during the dynamic, supersonic parachute inflation process: NASA's rover Curiosity
	data~\cite{cruz2014reconstruction}~(blue); second FSI simulation without the contribution to FSI of the suspension line subsystem (orange); 
	first FSI simulation with the contribution to FSI of the suspension line subsystem (green); and contribution of the suspension line subsystem 
	to total drag extracted from the first FSI simulation (red).}
  \label{fig:pid_drag}
\end{figure}

\Cref{fig:pid_drag} reports the time-histories of the total drag force predicted by both FSI simulations described above.
For reference, this figure also includes: the contribution of the suspension line subsystem to the total drag, for the first
FSI simulation where this contribution is accounted for; and the time-history of the total drag generated by the parachute system 
of NASA's rover Curiosity as measured during Mars landing~\cite{cruz2014reconstruction}. Regarding the first FSI simulation, the
reader can observe that the drag force generated by the suspension line subsystem is relatively negligible. From the total drag result
of the second FSI simulation however, the reader can infer that even though the suspension line subsystem is {\it directly} responsible
for a negligible part of the total drag, its effect on the total drag generated by the DGB parachute system is significant.
Indeed, the disturbances created ahead of the canopy by the shocks induced by this subsystem disrupt the bow shock of the overall 
DGB parachute system, mix the high pressure and low pressure flows, and as shown in \Cref{fig:pid_drag}, reduce the total drag force. 
Most importantly, the reader can also observe that the time-evolution of the total drag predicted by the first FSI simulation, which 
accounts for the effect of the suspension line subsystem on FSI, is in good agreement with the counterpart data measured during the Mars 
landing of Curiosity (relative error of less than $10\%$): this provides one successful validation of the master-slave kinematic approach 
proposed in this paper for modeling cable-driven FSI.

\section{Summary and conclusions}
\label{sec:conclusions}
In this paper, an embedded boundary approach for computing the Fluid-Structure Interaction (FSI) of cable systems or subsystems is presented.
In this approach, the dynamics of a solid cable are captured by a Finite Element (FE) semi-discretization of the cable centerline 
based on conventional beam or cable elements; however, the exact geometry of the cable is also represented, via a discretization of
the surface of the cable that is embedded in the fluid mesh. The embedded discrete surface is endowed with kinematics, and these
are slaved to the motion and deformation of the master beam/cable elements. The flow-induced loads are computed on the embedded discrete
surface and an energy-conserving method is applied to transfer these loads to the FE representation of the centerline of the cable.
The proposed approach for capturing cable-driven FSI is incorporated in an Eulerian framework for FSI equipped with adaptive mesh refinement
and applied to the solution of two highly nonlinear, turbulent FSI problems: an aerial refueling model problem; and a challenging, dynamic, supersonic parachute inflation problem
associated with the Mars landing of NASA's rover Curiosity. For the first problem, the proposed approach is shown to reproduce the same results as an alternative approach known as the ``dressing'' method.
For the second problem for which the dressing method is inapplicable, the proposed approach performs very well: it reproduces the total drag generated by the canopy and suspension lines
of the parachute system and measured in flight during the Mars landing of Curiosity.

\section*{Acknowledgments}

The authors acknowledge partial support by the Jet Propulsion Laboratory (JPL) under Contract JPL-RSA No. 1590208, and 
partial support by the National Aeronautics and Space Administration (NASA) under Early Stage Innovations (ESI) Grant NASA-NNX17AD02G. 
Daniel Huang also thanks Trond Kvamsdal for enlightening discussions on this topic.

\bibliography{myref}{}

\begin{thebibliography}{10}
\providecommand \doibase [0]{http://dx.doi.org/}%

\bibitem{fan2014simulation}
Fan Y, Xia J. Simulation of 3D parachute fluid--structure interaction based on
  nonlinear finite element method and preconditioning finite volume method.
  {\it Chinese Journal of Aeronautics} 2014\string; 27(6)\string: 1373--1383.

\bibitem{tezduyar2010space}
Tezduyar TE, Takizawa K, Moorman C, Wright S, Christopher J. Space--time finite
  element computation of complex fluid--structure interactions. {\it
  International Journal for Numerical Methods in Fluids} 2010\string;
  64(10-12)\string: 1201--1218.

\bibitem{huang2018simulation}
Huang Z, Avery P, Farhat C, Rabinovitch J, Derkevorkian A, Peterson LD.
  Simulation of parachute inflation dynamics using an Eulerian computational
  framework for fluid-structure interfaces evolving in high-speed turbulent
  flows. In: AIAA. ; 2018\string: 1540.

\bibitem{sengupta2009supersonic}
Sengupta A, Kelsch R, Roeder J, Wernet M, Witkowski A, Kandis M. Supersonic
  performance of disk-gap-band parachutes constrained to a 0-degree trim angle.
  {\it Journal of Spacecraft and Rockets} 2009\string; 46(6)\string:
  1155--1163.

\bibitem{gao2016numerical}
Gao X, Zhang Q, Tang Q. Numerical modelling of Mars supersonic disk-gap-band
  parachute inflation. {\it Advances in Space Research} 2016\string;
  57(11)\string: 2259--2272.

\bibitem{kim20093}
Kim Y, Peskin CS. 3-D parachute simulation by the immersed boundary method.
  {\it Computers \& Fluids} 2009\string; 38(6)\string: 1080--1090.

\bibitem{jaiman2009fully}
Jaiman RK, Shakib F, Oakley OH, Constantinides Y. Fully coupled fluid-structure
  interaction for offshore applications. In: American Society of Mechanical
  Engineers. ; 2009\string: 757--765.

\bibitem{holmes2006simulation}
Holmes S, Oakley OH, Constantinides Y. Simulation of riser VIV using fully
  three dimensional CFD simulations. In: American Society of Mechanical
  Engineers. ; 2006\string: 563--570.

\bibitem{herfjord1999assessment}
Herfjord K, Drange S, Kvamsdal T. Assessment of vortex-induced vibrations on
  deepwater risers by considering fluid-structure interaction. {\it Journal of
  Offshore Mechanics and Arctic Engineering} 1999\string; 121(4)\string:
  207--212.

\bibitem{lofthouse2017cfd}
Lofthouse AJ, Nathan B. CFD Modeling of B-52 and KC-135 in air refueling
  formation. In: AIAA. ; 2017\string: 4236.

\bibitem{ro2010modeling}
Ro K, Kamman JW. Modeling and simulation of hose-paradrogue aerial refueling
  systems. {\it Journal of Guidance, Control, and Dynamics} 2010\string;
  33(1)\string: 53--63.

\bibitem{zhu2007modeling}
Zhu Z, Meguid S. Modeling and simulation of aerial refueling by finite element
  method. {\it International Journal of Solids and Structures} 2007\string;
  44(24)\string: 8057--8073.

\bibitem{styuart2011numerical}
Styuart A, Gaston R, Yamashiro H, Stirling R, Mor M. Numerical simulation of
  hose whip phenomenon in aerial refueling. In: AIAA. ; 2011\string: 6211.

\bibitem{kim20062}
Kim Y, Peskin CS. 2--D parachute simulation by the immersed boundary method.
  {\it SIAM Journal on Scientific Computing} 2006\string; 28(6)\string:
  2294--2312.

\bibitem{Vienna}
Farhat C. Large-scale nonlinear aeroelastic computations: flutter, LCO and
  buffet investigations. In: WCCM V. ; 2002.

\bibitem{geuzaine2003aeroelastic}
Geuzaine P, Brown G, Harris C, Farhat C. Aeroelastic dynamic analysis of a full
  F-16 configuration for various flight conditions. {\it AIAA Journal}
  2003\string; 41(3)\string: 363--371.

\bibitem{wang2011algorithms}
Wang K, Rallu A, Gerbeau JF, Farhat C. Algorithms for interface treatment and
  load computation in embedded boundary methods for fluid and fluid--structure
  interaction problems. {\it International Journal for Numerical Methods in
  Fluids} 2011\string; 67(9)\string: 1175--1206.

\bibitem{farhat2012fiver}
Farhat C, Gerbeau JF, Rallu A. FIVER: A finite volume method based on exact
  two-phase Riemann problems and sparse grids for multi-material flows with
  large density jumps. {\it Journal of Computational Physics} 2012\string;
  231(19)\string: 6360--6379.

\bibitem{peskin1977numerical}
Peskin CS. Numerical analysis of blood flow in the heart. {\it Journal of
  Computational Physics} 1977\string; 25(3)\string: 220--252.

\bibitem{choi2007immersed}
Choi JI, Oberoi RC, Edwards JR, Rosati JA. An immersed boundary method for
  complex incompressible flows. {\it Journal of Computational Physics}
  2007\string; 224(2)\string: 757--784.

\bibitem{tseng2003ghost}
Tseng YH, Ferziger JH. A ghost-cell immersed boundary method for flow in
  complex geometry. {\it Journal of Computational Physics} 2003\string;
  192(2)\string: 593--623.

\bibitem{liu2006modified}
Liu T, Khoo B, Xie W. The modified ghost fluid method as applied to extreme
  fluid-structure interaction in the presence of cavitation. {\it Commun.
  Comput. Phys.} 2006\string; 1(5)\string: 898--919.

\bibitem{lakshminarayan2014embedded}
Lakshminarayan V, Farhat C, Main A. An embedded boundary framework for
  compressible turbulent flow and fluid--structure computations on structured
  and unstructured grids. {\it International Journal for Numerical Methods in
  Fluids} 2014\string; 76(6)\string: 366--395.

\bibitem{wang2015computational}
Wang K, Lea P, Farhat C. A computational framework for the simulation of
  high-speed multi-material fluid--structure interaction problems with dynamic
  fracture. {\it International Journal for Numerical Methods in Engineering}
  2015\string; 104(7)\string: 585--623.

\bibitem{main2017enhanced}
Main A, Zeng X, Avery P, Farhat C. An enhanced FIVER method for multi-material
  flow problems with second-order convergence rate. {\it Journal of
  Computational Physics} 2017\string; 329\string: 141--172.

\bibitem{huang2018family}
Huang DZ, De~Santis D, Farhat C. A family of position-and
  orientation-independent embedded boundary methods for viscous flow and
  fluid--structure interaction problems. {\it Journal of Computational Physics}
  2018\string; 365\string: 74--104.

\bibitem{borker2019mesh}
Borker R, Huang D, Grimberg S, Farhat C, Avery P, Rabinovitch J. Mesh
  adaptation framework for embedded boundary methods for computational fluid
  dynamics and fluid-structure interaction. {\it International Journal for
  Numerical Methods in Fluids} 2019.

\bibitem{hirt1974arbitrary}
Hirt CW, Amsden AA, Cook J. An arbitrary Lagrangian-Eulerian computing method
  for all flow speeds. {\it Journal of Computational Physics} 1974\string;
  14(3)\string: 227--253.

\bibitem{donea1982arbitrary}
Donea J, Giuliani S, Halleux JP. An arbitrary Lagrangian-Eulerian finite
  element method for transient dynamic fluid-structure interactions. {\it
  Computer Methods in Applied Mechanics and Engineering} 1982\string;
  33(1-3)\string: 689--723.

\bibitem{farhat2001discrete}
Farhat C, Geuzaine P, Grandmont C. The discrete geometric conservation law and
  the nonlinear stability of ALE schemes for the solution of flow problems on
  moving grids. {\it Journal of Computational Physics} 2001\string;
  174(2)\string: 669--694.

\bibitem{farhat1998load}
Farhat C, Lesoinne M, Le~Tallec P. Load and motion transfer algorithms for
  fluid/structure interaction problems with non-matching discrete interfaces:
  momentum and energy conservation, optimal discretization and application to
  aeroelasticity. {\it Computer Methods in Applied Mechanics and Engineering}
  1998\string; 157(1-2)\string: 95--114.

\bibitem{CMS}
CMSoft, Inc. MATCHER Version 2.0. {\it User's Reference Manual,} 2007.

\bibitem{farhat1995implicit}
Farhat C, Crivelli L, G{\'e}radin M. Implicit time integration of a class of
  constrained hybrid formulations part I: spectral stability theory. {\it
  Computer Methods in Applied Mechanics and Engineering} 1995\string;
  125(1-4)\string: 71--107.

\bibitem{roe1981approximate}
Roe PL. Approximate Riemann solvers, parameter vectors, and difference schemes.
  {\it Journal of Computational Physics} 1981\string; 43(2)\string: 357--372.

\bibitem{van1979towards}
Van~Leer B. Towards the ultimate conservative difference scheme. V. A
  second-order sequel to Godunov's method. {\it Journal of Computational
  Physics} 1979\string; 32(1)\string: 101--136.

\bibitem{laney1998computational}
Laney CB. {\it Computational gasdynamics}.
\newblock Cambridge university press .
\newblock 1998.

\bibitem{fedkiw1999non}
Fedkiw RP, Aslam T, Merriman B, Osher S. A non-oscillatory Eulerian approach to
  interfaces in multimaterial flows (the ghost fluid method). {\it Journal of
  Computational Physics} 1999\string; 152(2)\string: 457--492.

\bibitem{berger1989local}
Berger MJ, Colella P. Local adaptive mesh refinement for shock hydrodynamics.
  {\it Journal of Computational Physics} 1989\string; 82(1)\string: 64--84.

\bibitem{maubach1995local}
Maubach JM. Local bisection refinement for n-simplicial grids generated by
  reflection. {\it SIAM Journal on Scientific Computing} 1995\string;
  16(1)\string: 210--227.

\bibitem{stevenson2008completion}
Stevenson R. The completion of locally refined simplicial partitions created by
  bisection. {\it Mathematics of Computation} 2008\string; 77(261)\string:
  227--241.

\bibitem{vanella2014adaptive}
Vanella M, Posa A, Balaras E. Adaptive mesh refinement for immersed boundary
  methods. {\it Journal of Fluids Engineering} 2014\string; 136(4)\string:
  040909.

\bibitem{roma1999adaptive}
Roma AM, Peskin CS, Berger MJ. An adaptive version of the immersed boundary
  method. {\it Journal of Computational Physics} 1999\string; 153(2)\string:
  509--534.

\bibitem{spalart1992one}
Spalart P, Allmaras S. A one-equation turbulence model for aerodynamic flows.
  In: AIAA. ; 1992\string: 439.

\bibitem{farhat2010robust}
Farhat C, Rallu A, Wang K, Belytschko T. Robust and provably second-order
  explicit--explicit and implicit--explicit staggered time-integrators for
  highly non-linear compressible fluid--structure interaction problems. {\it
  International Journal for Numerical Methods in Engineering} 2010\string;
  84(1)\string: 73--107.

\bibitem{cruz2014reconstruction}
Cruz JR, Way DW, Shidner JD, Davis JL, Adams DS, Kipp DM. Reconstruction of the
  Mars science laboratory parachute performance. {\it Journal of Spacecraft and
  Rockets} 2014\string; 51(4)\string: 1185--1196.

\bibitem{lin2010flexible}
Lin JK, Shook LS, Ware JS, Welch JV. Flexible material systems testing. NASA
  Report CR-2010-216854;  2010.

\bibitem{vreman2004eddy}
Vreman A. An eddy-viscosity subgrid-scale model for turbulent shear flow:
  algebraic theory and applications. {\it Physics of Fluids} 2004\string;
  16(10)\string: 3670--3681.

\bibitem{militello1991first}
Militello C, Felippa CA. The first ANDES elements: 9-dof plate bending
  triangles. {\it Computer Methods in Applied Mechanics and Engineering}
  1991\string; 93(2)\string: 217--246.

\bibitem{mitchell1988unified}
Mitchell WF. {\it Unified multilevel adaptive finite element methods for
  elliptic problems}. PhD thesis. University of Illinois at Urbana-Champaign, ;
   1988.

\bibitem{huang2019homogenized}
Huang DZ, Wong ML, Lele SK, Farhat C. A homogenized flux-body force approach
  for modeling porous wall boundary conditions in compressible viscous flows.
  {\it arXiv preprint arXiv:1907.09632} 2019.

\end{thebibliography}

\end{document}